\documentclass[conference,compsoc]{IEEEtran}

\usepackage{cite}
\usepackage{amsmath,amssymb,amsfonts}
\usepackage{algorithmic}
\usepackage{graphicx}
\usepackage{textcomp}
\usepackage[table]{xcolor}   
\usepackage{booktabs}
\usepackage{threeparttable}
\usepackage{tabularx}
\usepackage{array}
\usepackage{pgffor}
\usepackage{subcaption}
\usepackage{multirow}
\usepackage[normalem]{ulem}
\usepackage{paralist}
\usepackage{rotating}
\usepackage{tcolorbox}
\usepackage{tikz}
\usepackage{url}
\usepackage[hidelinks]
{hyperref}  
\usepackage{orcidlink}  
\usepackage{float}
\usepackage{placeins}
\usepackage[available,functional,reproduced]{ieeebadges} 

\newcommand{\best}[1]{\cellcolor{green!20}{#1}}
\newcommand{\worst}[1]{\cellcolor{red!20}{#1}}
\newcolumntype{L}[1]{>{\raggedright\arraybackslash}p{#1}}

\renewcommand{\paragraph}[1]{\textbf{#1.}}

\IEEEoverridecommandlockouts

\begin{document}

\title{Quiet Triggers, Loud Footprints: A Tri-Space Measurement Study of Backdoor Stealthiness}

\author{\IEEEauthorblockN{Behrad Tajalli$^{1}$\,\orcidlink{0009-0001-0636-3385},
Henk Berendsen$^{1}$\,\orcidlink{0009-0007-6902-8844},
Xiaoyun Xu$^{1*}$\,\orcidlink{0000-0001-5726-163X}\thanks{$^*$Corresponding author: Xiaoyun Xu.}, and
Stjepan Picek$^{2,1}$\,\orcidlink{0000-0001-7509-4337}}
\IEEEauthorblockA{$^{1}$Radboud University, Nijmegen, The Netherlands}
\IEEEauthorblockA{$^{2}$University of Zagreb Faculty of Electrical Engineering and Computing, Zagreb, Croatia}}

\maketitle

\begin{abstract}
Backdoor attacks on deep neural networks are commonly evaluated by attack success rate and benign accuracy, yet their stealthiness remains inconsistently defined and measured. Existing work often studies only one observable aspect of stealthiness, such as trigger imperceptibility, feature-space separability, parameter-space concentration, or evasion of particular defenses. This creates a fragmented picture: an attack may appear stealthy under one metric while leaving conspicuous footprints elsewhere.
This paper presents a tri-space measurement study of backdoor stealthiness in image classification. We distinguish latent stealthiness (the ideal but unobservable difficulty of distinguishing a backdoored system from a benign one) from measurable footprints in three spaces: input, feature, and parameter space. Under this framework, we evaluate representative backdoor attacks across four datasets and three model architectures using a broad suite of footprint metrics.

Our results show that stealthiness is not monolithic: attacks that are quiet in one space are often loud in another, and existing attacks rarely achieve consistently small footprints across all three spaces.
We further analyze agreement and disagreement among stealthiness metrics and study how footprint measurements interact with attack success rate. Based on observed blind spots in existing measurements, we introduce two diagnostic metrics: Class-Specific Davies--Bouldin Index (CDBI), which captures class-conditional feature overlap, and TAC--UCLC Product (TUP), which identifies channels that are both trigger-responsive and downstream influential. These metrics are intended to augment, not replace, the broader measurement suite. Finally, we provide practical guidance for future stealthiness evaluations. Our findings suggest that claims of backdoor stealthiness should be supported by multi-space footprint measurements rather than by a single metric, defense, or visual inspection.\end{abstract}


\IEEEpeerreviewmaketitle

\section{Introduction}
\label{sec:introduction}

Deep neural networks (DNNs) have demonstrated prominent success across computer vision tasks, but they remain vulnerable to backdoor attacks. In a backdoor attack, an adversary implants hidden behavior into a model: the model behaves normally on clean inputs, but inputs containing an attacker-chosen trigger are misclassified to a target class. Backdoors can be introduced through poisoned training data, modified training procedures, or direct manipulation of model weights.

A central objective of backdoor attacks is stealthiness. Yet stealthiness is used inconsistently across the literature. Some attacks emphasize visually imperceptible triggers; others attempt to hide poisoned samples in feature space, distribute the backdoor across parameters, or evade a particular defense. These are related but distinct claims. An attack can be visually quiet while leaving a large feature-space footprint, or it can appear parameter-quiet while requiring an input trigger that is easy to detect.

\paragraph{Motivation} This paper treats backdoor stealthiness as a measurement problem. We distinguish latent stealthiness (the ideal but unobservable difficulty of distinguishing a backdoored system from a benign one) from measurable footprints in three spaces: input, feature, and parameter space. Input-space footprints measure how much the trigger changes the image. Feature-space footprints measure whether triggered and benign latent representations are separated inside the model. Parameter-space footprints measure whether backdoor behavior is concentrated in identifiable neurons, channels, or weights.

Our goal is not to produce a single leaderboard of ``the most stealthy'' attack. Backdoor attacks differ in threat model, poisoning mechanism, trigger construction, and training access. Instead, we provide a controlled measurement study: we evaluate representative attacks under a common protocol and ask where each attack leaves observable footprints. This framing also avoids defining stealthiness by defense evasion. Defense outcomes are useful external signals, but defenses are incomplete and evolve over time; evading a defense does not imply the absence of a footprint.

Unlike prior work that introduces a new attack, defense, or single stealthiness score, our contribution is a measurement result: commonly used stealthiness proxies do not behave as interchangeable estimates of one latent property. The same attack can have a small footprint in one observation space and a large footprint in another. This empirical non-substitutability is the reason we organize our analysis around spaces rather than around a single overall stealthiness ranking.

\paragraph{Our contribution} We explain the background knowledge required to understand our paper in Section~\ref{sec:background}. In Section~\ref{sec:benchmark_results}, we evaluate ten representative attacks across four datasets and three model architectures using a broad suite of input-, feature-, and parameter-space metrics. 

Motivated by the observed failure modes of these metrics, we introduce two diagnostic additions in Section~\ref{sec:diagnostic_metrics}. The Class-specific Davies--Bouldin Index (CDBI) measures class-conditional feature overlap and helps interpret cases where global cluster scores are affected by class fragmentation or imbalance. The TAC--UCLC Product (TUP) identifies channels that are both trigger-responsive and downstream influential. These metrics are intended to augment the measurement suite, not replace it.

In Section~\ref{sec:metric_disagreement}, we find that stealthiness is not monolithic. Metrics agree strongly in some local cases, such as closely related input-space distortions, but disagree across spaces. Attacks that appear quiet in one space are often loud in another. This suggests that stealthiness claims should be supported by multi-space evidence rather than by a single metric, defense, visual inspection, or aggregate score.

We also perform external checks on footprint measurements in Section~\ref{sec:robustness_external_checks}. In Section \ref{sec:asr_effectiveness}, we study how footprint measurements interact with attack success rate, showing why low-ASR configurations should be interpreted as a separate effectiveness regime rather than simply discarded. The backdoor defense analysis in Sections~\ref{sec:defense_external_checks} and \ref{sec:defense_association} shows that defenses can exploit some measured footprints, but do not provide a universal proxy for stealthiness.

In Section~\ref{sec:guidance_limitations}, we explain the code artifact we used to perform our evaluation, the limitations of this evaluation and provide some recommendations for future work on backdoor stealthiness. Then, we discuss related work in Section~\ref{sec:related_work} and finally conclude the paper in Section~\ref{sec:conclusion}.
\section{Background}
\label{sec:background}

\subsection{Latent Stealthiness and Observable Footprints}
\label{sec:latent_footprints}

We use latent stealthiness to denote the ideal property that the presence of a backdoor is difficult to distinguish from benign behavior under relevant inspections. This property is not directly observable: no finite set of metrics, visual checks, or defenses can exhaust all possible detection strategies.

We therefore operationalize stealthiness through observable footprints. A footprint is a measurable difference between clean and backdoored behavior in a particular space. We study three such spaces (Figure~\ref{fig:threat_model} in Appendix~\ref{app:threat}).

\paragraph{Input space} Input-space metrics compare a benign image $x$ with its triggered counterpart $T(x)$. They capture visual or perceptual changes introduced by the trigger, such as pixel-level distortion, structural similarity, perceptual distance, or semantic change under a reference classifier.

\paragraph{Feature space} Feature-space metrics compare representations extracted by the backdoored model, typically from the penultimate layer. They capture whether triggered samples form separable clusters or distributional shifts relative to benign samples.

\paragraph{Parameter space} Parameter-space metrics inspect the model internals. They capture whether the backdoor is concentrated in neurons, channels, or weights that are unusually sensitive to the trigger or unusually influential downstream.

Feature-space and parameter-space measurements are related but not redundant. Feature metrics are data-dependent: they ask whether triggered inputs hide among benign representations for a given sample distribution. Parameter metrics are mechanism-dependent: they ask whether the learned backdoor is concentrated in identifiable internal components. A model can produce similar feature distributions while relying on different internal mechanisms, and parameter-level anomalies may not be visible under a particular feature sample.

\subsection{Footprint Metrics}
\label{sec:footprint_metrics}

We group metrics by the footprint they measure in Table~\ref{tab:metrics}. In the metric-agreement analysis (Section~\ref{sec:metric_disagreement}), all metrics are oriented so that larger values indicate larger footprints; the benchmark tables in Section~\ref{sec:benchmark_results} report each metric in its native orientation (indicated by arrows), and the residual association analysis (Figure~\ref{fig:residual_correlations}) defines its own alignment in the caption.
Definitions and implementation details for standard footprint metrics are provided in the supplementary material and in the artifact; the main text defines CDBI and TUP because they are diagnostic additions introduced in this work.

\begin{table}[htbp]
\centering
\scriptsize 
\caption{Stealthiness footprint metrics used in our benchmark.}
\label{tab:metrics}
\begin{tabularx}{\linewidth}{@{} l l >{\raggedright\arraybackslash}X @{}}
\toprule
Space & Metrics & Captures \\
\midrule
Input & $\ell_1$, $\ell_2$, $\ell_\infty$, MSE, PSNR & Pixel-level distortion \\
Input & SSIM, LPIPS, pHash, IS, SAM & Structural/perceptual/semantic change \\
Feature & SS, DSWD & Cluster separability and distribution shift \\
Feature & CDBI & Class-conditional feature overlap \\
Parameter & UCLC, TAC & Channel influence and trigger response \\
Parameter & TUP & Joint trigger response and influence \\
\bottomrule
\end{tabularx}
\end{table}

\section{Benchmark Scope and Results}
\label{sec:benchmark_results}

\paragraph{Attack families}
The evaluated attacks cover five broad families. BadNets~\cite{gu2019badnets} and Blend~\cite{chen2017blend} represent patch/blend poisoning attacks with simple, fixed triggers. WaNet~\cite{nguyen2021wanet} and Bpp\-Attack~\cite{wang2022bppattack} represent imperceptible input-space attacks: WaNet uses image warping, while BppAttack uses image quantization to reduce visible perturbations. Adap-Patch and Adap-Blend~\cite{qi2023adap} are adaptive attacks designed to reduce feature-space separability. Narcissus~\cite{zeng2022narcissus} and Grond~\cite{xu2025grond} are clean-label attacks with stronger poisoning constraints; Grond additionally optimizes for parameter-space behavior. DFST~\cite{cheng2021dfst} and DFBA~\cite{cao2024dfba} represent stronger supply-chain or model-modification settings, where the attacker has more control over the training or model-modification process. These families are not directly comparable as an attack-strength leaderboard, but they are useful for measuring where different design objectives leave observable footprints.

\paragraph{Datasets and Architectures} The datasets (CIFAR-10, CIFAR-100, Tiny-ImageNet, and Imagenette) span small natural-image benchmarks and higher-resolution settings, while the architecture set includes two convolutional models (ResNet18 and VGG) and a transformer (ViT-Small). This design lets us test whether footprint conclusions are attack-family- or architecture-specific, while keeping the benchmark controlled enough for per-metric comparison. We use ResNet18 for the primary tri-space comparison because all ten
attacks are available under this architecture. Appendix~\ref{sec:experimental_settings} and the supplementary material report the experimental settings used to instantiate this benchmark, including dataset normalization statistics, training settings, attack-specific hyperparameters, and metric implementation details.

\paragraph{Motivation}
The benchmark is designed to compare observable footprints rather than to equalize all threat models. The considered backdoor attacks differ in attacker access, poisoning constraints, trigger construction, and optimization objective. We therefore avoid interpreting the results as an absolute attack leaderboard. Instead, we use the shared protocol to ask a more specific question: when an attack is evaluated under its intended setting, where does it leave measurable evidence? This distinction is important because an attack with stronger access may reduce one footprint while leaving another, and an attack with weaker access may appear quiet only because it fails to implant a high-ASR backdoor. The benchmark should therefore be read as a map of input-, feature-, and parameter-space footprints under controlled evaluation, not as a single ranking of attack strength.

\paragraph{Results Overview} Table~\ref{tab:attack-performance-4datasets} reports the BA and ASR values for the ResNet18 models used in the main tri-space analysis.

\renewcommand{\arraystretch}{0.8}
\begin{table}[t]
    \centering
    \scriptsize
    \setlength\tabcolsep{3pt}
    \caption{Backdoor performance of ResNet18 models. The benign accuracy (BA) is shown for both benign and backdoored models. The poisoning rate (PR) is specified for backdoored models, and their attack success rate (ASR) is also reported. For the backdoored models, ASR scores below 60\% are highlighted in red.
    *The poisoning rates vary per dataset.}
    \label{tab:attack-performance-4datasets}
    \begin{tabular}{lccccccccc}
    \toprule
        \multirow{2}{*}{Attack} & \multirow{2}{*}{PR (\%)} & \multicolumn{2}{c}{C10} & \multicolumn{2}{c}{C100} & \multicolumn{2}{c}{Tiny} & \multicolumn{2}{c}{Imagenette}\\
         \cmidrule(lr){3-4}
         \cmidrule(lr){5-6}
         \cmidrule(lr){7-8}
         \cmidrule(lr){9-10}
         ~ & ~ & BA & ASR & BA & ASR & BA & ASR & BA & ASR \\
         \midrule
         Benign & - & 94.7 & - & 73.3 & - & 59.21 & - & 87.7 & -\\
         \cmidrule(lr){2-10}
         \multirow{2}{*}{BadNets} & 5 & 94.5 & 100 & 72.9 & 100 & 58.37 & 99.97 & 88.3 & 100\\
         ~ & 0.3 & 94.7 & 100 & 73.5 & 99.8 & 59.12 & 94.04 & 88.1 & 86.7\\
         \cmidrule(lr){2-10}
         \multirow{2}{*}{Blend} & 5 & 94.3 & 99.8 & 72.4 & 99.7 & 58.24 & 99.38 & 88.6 & 98.9\\
         ~ & 0.3 & 94.8 & 94.8 & 73.6 & 88.6 & 58.74 & 84.22 & 88.6 & 73.5\\
         \cmidrule(lr){2-10}
         \multirow{2}{*}{WaNet} & 5 & 93.4 & 98.5 & 66.8 & 99.8 & 57.88 & 99.93 & 86.6 & 97.2\\
         ~ & \scriptsize{2/0.3/0.3/0.3$^*$} & 93.5 & 95.3 & 66.6 & 91.7 & 58.64 & 87.94 & 79.1 & 61.0\\
         \cmidrule(lr){2-10}
         \multirow{2}{*}{Bpp} & 5 & 94.4 & 99.8 & 73.0 & 99.9 & 59.36 & 98.43 & 88.9 & 73.6 \\
         ~ & \scriptsize{2/0.3/0.3/0.3$^*$} & 94.6 & 69.5 & 73.0 & 88.0 & 58.56 & \textcolor{red}{4.72} & 88.3 & \textcolor{red}{1.02} \\
         \cmidrule(lr){2-10}
         \multirow{2}{1cm}{Adap\\Patch} & 5 & 93.9 & \textcolor{red}{30.1} & 70.9 & 66.7 & 56.97 & \textcolor{red}{41.62} & 87.6 & \textcolor{red}{20.2}\\
         ~ & 0.3 & 94.5 & 92.2 & 72.7 & \textcolor{red}{2.98} & 58.39 & \textcolor{red}{0.26} & 88.1 & \textcolor{red}{0.565}\\
         \cmidrule(lr){2-10}
         \multirow{2}{1cm}{Adap\\Blend} & 5 & 94.0 & \textcolor{red}{53.9} & 71.8 & 90.6 & 57.59 & 96.36 & 87.6 & 94.9\\
         ~ & 0.3 & 94.6 & 87.9 & 72.6 & 82.0 & 58.9 & 81.58 & 88.5 & \textcolor{red}{54.4} \\
         \cmidrule(lr){2-10}
         \multirow{2}{*}{DFST} & 5 & 95.0 & 100 & 73.6 & 100 & 59.23 & 100 & 88.7 & 99.5\\
         ~ & 0.3 & 94.8 & 100 & 73.6 & 99.9 & 58.79 & 99.84 & 88.4 & 99.7\\
         \cmidrule(lr){2-10}
         \multirow{2}{*}{Narcissus} & \scriptsize{5/0.7/0.4/5$^*$} & 94.4 & 100 & 73.0 & 70.9 & 59.04 & 98.82 & 87.8 & 86.1\\
         ~ & 0.3 & 94.7 & 98.9 & 72.8 & \textcolor{red}{33.1} & 58.97 & 98.97 & 88.7 & \textcolor{red}{13.6} \\
         \cmidrule(lr){2-10}
         \multirow{2}{*}{Grond} & \scriptsize{5/0.7/0.4/5$^*$} & 94.7 & 92.9 & 74.7 & \textcolor{red}{58.7} & 59.02 & 76.32 & 87.0 & 60.2\\
         ~ & 0.3 & 94.5 & 76.6 & 74.7 & \textcolor{red}{43.5} & 58.68 & 74.29 & 87.4 & \textcolor{red}{1.81}\\
         \cmidrule(lr){2-10}
         DFBA & - & 94.3 & 100 & 72.0 & 100 & 55.95 & 100 & 87.6 & 100\\
         \bottomrule
    \end{tabular}
\end{table}



The tri-space comparison in Table~\ref{tab:trispace_summary} should be read as a footprint map rather than a leaderboard. Each attack family is optimized under different constraints, so the relevant question is not which family is universally stealthy, but where its observable footprint appears. The summary reveals three recurring patterns. First, attacks designed to reduce input perturbations, such as imperceptible or warping-based triggers, can still leave feature- or parameter-space evidence. Second, attacks designed to reduce feature separability can have unstable ASR or non-negligible input footprints. Third, attacks with small parameter-space footprints often achieve this under stronger constraints or stronger attacker access. These patterns motivate the metric-disagreement analysis in Section~\ref{sec:metric_disagreement}: if spaces capture different footprints, then a single stealthiness metric cannot support a general stealthiness claim. The per-space footprints of the evaluated attacks on ResNet18 are discussed in the paragraphs below.

\begin{table}[t]
\centering
\caption{Qualitative tri-space footprint summary. Entries summarize whether each attack family tends to have smaller, mixed, or larger measured footprints in each space. The summary is not a leaderboard; it identifies where observable footprints tend to appear.}
\label{tab:trispace_summary}
\begin{tabular}{llll}
\toprule
Attack family & Input & Feature & Parameter \\
\midrule
Patch/blend & Low--mixed & High & High \\
Imperceptible & Low & Mixed & Mixed--high \\
Adaptive & Mixed & Low & Mixed \\
Clean-label & High & Low--mixed & Low \\
Supply-chain & Mixed & High & Mixed--high \\
\bottomrule
\end{tabular}
\end{table}

\paragraph{Input-space footprints}
Table~\ref{tab:input-stealthiness-all} reports the input-space footprint measurements for ResNet18 models. In this and the following benchmark tables (Tables~\ref{tab:input-stealthiness-all}, \ref{tab:results_feature_resnet18}, and~\ref{tab:uclc-tac}), highlighted extrema identify the smallest and largest measured footprints under the evaluated configurations; because attacks differ in access and design constraints, they should not be interpreted as proof that one attack is categorically superior to another. Input-space metrics generally agree that BppAttack and WaNet leave small visual/perceptual footprints, while Narcissus and DFST often leave larger perturbations. However, individual input metrics disagree on localized triggers: $\ell_1$ and $\ell_2$ can favor small patch triggers, whereas $\ell_\infty$ penalizes them because a few pixels change strongly. This illustrates why input-space stealthiness should not be inferred from a single distortion metric.

The input-space table also illustrates a difference between attacks that are visually small by construction and attacks that satisfy other constraints. BppAttack and WaNet obtain small footprints across many input metrics because their triggers are distributed or quantization-based rather than large localized patches. By contrast, Narcissus and Grond operate under clean-label constraints and must preserve the original label during poisoning, which changes the optimization trade-off. The adaptive attacks and Narcissus are especially asymmetric: their training-time perturbations are smaller than their inference-time triggers, so input-space conclusions depend on whether the evaluator measures the poisoned training samples or the backdoored inference samples. DFBA shows the opposite scale effect: because its trigger size is fixed, its relative input-space footprint is smaller on larger images. These cases show why input-space footprint claims should report the trigger setting, dataset resolution, and whether train-time or test-time triggers are being measured.

\begin{table*}[t]
    \scriptsize
    \centering
    \caption{Input-space footprint measurements on four datasets.}
    \renewcommand{\arraystretch}{0.7}
    \label{tab:input-stealthiness-all}
    \begin{tabular}{llcccccccccc}
    \toprule
    \multirow{2}{*}{Dataset} & Metric & $l_1 \downarrow$ & $l_2 \downarrow$ & $l_\infty \downarrow$ & MSE $\downarrow$& PSNR $\uparrow$& SSIM $\uparrow$& LPIPS $\downarrow$& IS $\downarrow$& pHash $\uparrow$& SAM $\downarrow$\\
    ~ & Ideal value & 0 & 0 & 0 & 0 & $\infty$ & 1 & 0 & 0 & 100 & 0 \\
    \midrule
    \multirow{13}{*}{CIFAR-10} & BadNets & \best{13.2} & 2.84 & 0.811 & 2.64e-3 & 25.8 & 0.956 & 4.41e-3 & 0.170 & 95.2 & \best{1.44e-3} \\
    ~ & Blend & 230 & 4.92 & 0.191 & 8.01e-3 & 21.0 & 0.784 & 0.0483 & 2.37 & 89.8 & 0.0830 \\
    ~ & WaNet & 46.5 & 1.42 & 0.173 & 7.17e-4 & 31.5 & 0.978 & 4.32e-3 & 0.270 & 96.1 & 0.0119 \\
    ~ & BppAttack & 24.6 & \best{0.525} & \best{0.0196} & \best{8.97e-5} & \best{40.5} & \best{0.992} & \best{1.94e-4} & \best{0.150} & \best{99.2} & 0.0239 \\
    ~ & Adap-Patch (train) & 22.5 & 2.06 & 0.315 & 1.53e-3 & 28.7 & 0.972 & 4.78e-3 & 0.296 & 95.2 & 9.43e-3 \\
    ~ & Adap-Patch (test) & 74.1 & 6.77 & \worst{0.913} & 0.0151 & 18.2 & 0.956 & 0.0120 & 0.724 & 87.7 & 8.94e-3 \\
    ~ & Adap-Blend (train) & 87.0 & 2.63 & 0.142 & 2.30e-3 & 26.4 & 0.903 & 0.0136 & 0.823 & 92.4 & 0.0336 \\
    ~ & Adap-Blend (test) & 232 & 4.97 & 0.192 & 8.15e-3 & 20.9 & 0.783 & 0.0485 & 2.37 & 89.8 & 0.0831 \\
    ~ & DFST & 487 & \worst{10.8} & 0.489 & \worst{0.0400} & \worst{14.0} & 0.719 & 0.165 & 2.22 & 88.6 & 0.316 \\
    ~ & Narcissus (train) & 188 & 3.43 & 0.0627 & 3.83e-3 & 24.2 & 0.818 & 0.0545 & 1.44 & 91.8 & 0.101 \\
    ~ & Narcissus (test) & \worst{542} & 9.98 & 0.188 & 0.0325 & 14.9 & \worst{0.532} & \worst{0.184} & \worst{3.68} & \worst{82.2} & \worst{0.331} \\
    ~ & Grond & 87.6 & 1.62 & 0.0348 & 8.59e-4 & 30.7 & 0.934 & 2.92e-3 & 0.918 & 97.9 & 0.0833 \\
    ~ & DFBA & 13.5 & 2.99 & 0.844 & 2.94e-3 & 25.3 & \best{0.992} & 1.66e-3 & 0.159 & 97.4 & 2.83e-3 \\
    \midrule
    \multirow{13}{*}{CIFAR-100} & BadNets & \best{13.3} & 2.91 & 0.839 & 2.79e-3 & 25.6 & 0.955 & 5.39e-3 & 0.182 & 94.1 & \best{1.66e-3} \\
    ~ & Blend & 232 & 4.98 & 0.191 & 8.23e-3 & 20.8 & 0.765 & 0.0512 & \worst{2.25} & 89.0 & 0.0876 \\
    ~ & WaNet & 55.3 & 1.64 & 0.177 & 9.55e-4 & 30.2 & 0.968 & 6.72e-3 & 0.308 & 95.2 & 0.0154 \\
    ~ & BppAttack & 24.3 & \best{0.520} & \best{0.0196} & \best{8.83e-5} & \best{40.5} & 0.991 & \best{2.06e-4} & \best{0.138} & \best{99.1} & 0.0235 \\
    ~ & Adap-Patch (train) & 22.7 & 2.09 & 0.320 & 1.59e-3 & 28.5 & 0.971 & 4.78e-3 & 0.297 & 94.4 & 9.51e-3 \\
    ~ & Adap-Patch (test) & 74.5 & 6.96 & \worst{0.928} & 0.0159 & 18.0 & 0.956 & 0.0135 & 0.805 & 86.2 & 9.73e-3 \\
    ~ & Adap-Blend (train) & 87.9 & 2.66 & 0.142 & 2.36e-3 & 26.3 & 0.894 & 0.0144 & 0.796 & 91.5 & 0.0356 \\
    ~ & Adap-Blend (test) & 234 & 5.02 & 0.192 & 8.36e-3 & 20.8 & 0.764 & 0.0513 & \worst{2.25} & 89.0 & 0.0877 \\
    ~ & DFST & 509 & \worst{11.3} & 0.495 & \worst{0.0444} & \worst{13.5} & 0.708 & \worst{0.169} & 2.16 & 87.7 & 0.312 \\
    ~ & Narcissus (train) & 174 & 3.27 & 0.0627 & 3.50e-3 & 24.6 & 0.884 & 0.0147 & 0.622 & 94.3 & 0.126 \\
    ~ & Narcissus (test) & \worst{528} & 9.82 & 0.188 & 0.0315 & 15.0 & \worst{0.630} & 0.166 & 2.22 & \worst{84.6} & \worst{0.347} \\
    ~ & Grond & 86.9 & 1.61 & 0.0347 & 8.50e-4 & 30.7 & 0.929 & 4.97e-3 & 0.876 & 97.9 & 0.0720 \\
    ~ & DFBA & 13.5 & 3.02 & 0.843 & 3.04e-3 & 25.2 & \best{0.993} & 7.74e-3 & 0.146 & 95.9 & 4.18e-3 \\
    \midrule
    \multirow{13}{*}{Tiny-ImageNet} & BadNets & 58.36 & 4.99 & 0.73 & 2.23e-3 & 26.53 & 0.96 & 4.15e-2 & 0.11 & 91.79 & 3.47e-3\\
    ~ & Blend & 921.81 & 9.95 & 0.2 & 8.22e-3 & 20.85 & 0.81 & 0.11 & 1.11 & 89.44 & 0.11\\
    ~ & WaNet & 317.02 & 4.94 & 0.35 & 2.20e-3 & 26.58 & 0.93 & 4.03e-2 & 0.27 & 97.35 & 2.60e-2\\
    ~ & BppAttack & 96.26 & \best{1.04} & \best{1.96e-2} & \best{8.76e-5} & \best{40.57} & 0.99 & \best{8.75e-4} & 9.75e-2 & 99.42 & 3.14e-2\\
    ~ & Adap-Patch (train) & 23.54 & 2.17 & 0.33 & 4.22e-4 & 34.22 & 0.99 & 2.56e-3 & 5.07e-2 & 96.93 & 2.93e-3\\
    ~ & Adap-Patch (test) & 75.68 & 7.06 & \worst{0.96} & 4.11e-3 & 23.87 & 0.99 & 8.12e-3 & 0.17 & 92.05 & 3.05e-3\\
    ~ & Adap-Blend (train) & 358.75 & 5.45 & 0.15 & 2.47e-3 & 26.07 & 0.91 & 4.19e-2 & 0.49 & 91.8 & 4.67e-2\\
    ~ & Adap-Blend (test) & 955.04 & 10.28 & 0.2 & 8.77e-3 & 20.58 & 0.79 & 0.13 & 1.42 & 89.21 & 0.11\\
    ~ & DFST & 1.97e+3 & 22.16 & 0.53 & 4.18e-2 & 13.83 & 0.68 & 0.32 & 1.72 & 87.87 & 0.31\\
    ~ & Narcissus (train) & 1.23e+3 & 12.1 & 0.2 & 1.19e-2 & 19.23 & 0.65 & 0.22 & 1.55 & 85.56 & 0.26\\
    ~ & Narcissus (test) & \worst{3.29e+3} & \worst{33.31} & 0.59 & \worst{9.05e-2} & \worst{10.43} & \worst{0.36} & \worst{0.56} & \worst{4.73} & \worst{70.98} & \worst{0.5}\\
    ~ & Grond & 671.95 & 6.31 & 6.67e-2 & 3.25e-3 & 24.89 & 0.8 & 0.16 & 1.68 & 96.83 & 0.17\\
    ~ & DFBA & \best{13.34} & 3.01 & 0.87 & 7.49e-4 & 31.26 & \best{0.99} & 2.09e-3 & \best{4.42e-2} & \best{99.59} & \best{1.34e-3}\\
    \midrule
    \multirow{13}{*}{Imagenette} & BadNets & 73.3 & 6.94 & 0.879 & 2.54e-3 & 26.0 & 0.981 & 0.0152 & 0.0331 & 97.2 & 1.44e-3 \\
    ~ & Blend & 1.62e3 & 13.8 & 0.198 & 0.0100 & 20.0 & 0.697 & 0.203 & 3.10 & 89.9 & 0.0968 \\
    ~ & WaNet & 225 & 2.94 & 0.190 & 4.94e-4 & 33.3 & 0.981 & 0.0140 & 0.0775 & 98.0 & 0.0111 \\
    ~ & BppAttack & 151 & \best{1.30} & \best{0.0196} & \best{8.79e-5} & \best{40.6} & 0.983 & \best{1.87e-3} & 0.0990 & 99.4 & 0.0294 \\
    ~ & Adap-Patch (train) & 136 & 5.18 & 0.342 & 1.54e-3 & 28.2 & 0.967 & 0.0229 & 0.175 & 94.7 & 0.0101 \\
    ~ & Adap-Patch (test) & 424 & 16.7 & \worst{0.960} & 0.0147 & 18.4 & 0.955 & 0.0540 & 0.238 & 86.9 & 9.00e-3 \\
    ~ & Adap-Blend (train) & 615 & 7.35 & 0.147 & 2.87e-3 & 25.4 & 0.865 & 0.0769 & 0.844 & 92.0 & 0.0413 \\
    ~ & Adap-Blend (test) & 1.64e3 & 13.9 & 0.198 & 0.0102 & 19.9 & 0.697 & 0.203 & 3.09 & 89.9 & 0.0969 \\
    ~ & DFST & 3.22e3 & \worst{28.9} & 0.532 & \worst{0.0456} & \worst{13.5} & 0.692 & 0.335 & 1.52 & 88.1 & \worst{0.334} \\
    ~ & Narcissus (train) & 1.17e3 & 8.55 & 0.0627 & 3.81e-3 & 24.2 & 0.861 & 0.113 & 0.669 & 93.4 & 0.139 \\
    ~ & Narcissus (test) & \worst{3.31e3} & 24.6 & 0.188 & 0.0317 & 15.0 & \worst{0.615} & \worst{0.428} & \worst{3.37} & \worst{82.8} & 0.268 \\
    ~ & Grond & 476 & 3.66 & 0.0352 & 7.00e-4 & 31.6 & 0.882 & 0.0482 & 1.02 & 99.1 & 0.0521 \\
    ~ & DFBA & \best{13.5} & 3.05 & 0.855 & 4.90e-4 & 33.1 & \best{0.999} & 2.07e-3 & \best{0.0312} & \best{99.8} & \best{8.93e-4} \\
    \bottomrule
    \end{tabular}
\end{table*}

\paragraph{Feature-space footprints}
Table~\ref{tab:results_feature_resnet18} reports the feature-space footprint measurements for ResNet18 models. Because the data-free DFBA poisons no training samples, SS (computed over poisoned training data) is undefined for it and marked ``--'', while DSWD (computed from triggered test inputs) remains measurable.
Feature-space metrics show that adaptive and clean-label attacks can reduce global clean/poisoned separability, but this conclusion is sensitive to class fragmentation and attack success. SS and DSWD often rank attacks differently because SS measures cluster geometry in an embedding, while DSWD measures distributional shift along discriminative directions. Section~\ref{sec:cdbi_case} analyzes cases where CDBI provides additional diagnostic information.

\renewcommand{\arraystretch}{0.7}
\begin{table}[t]
\scriptsize
\centering
\setlength\tabcolsep{0.75pt}
\caption{Feature-space footprint of backdoored ResNet18 models trained on four datasets. The metrics used to evaluate each model are the silhouette score (SS) and discriminant sliced-Wasserstein distance (DSWD). For each metric and dataset, the \best{best} and \worst{worst} results are highlighted. Results that may be affected by the low ASR of the corresponding models (see Table~\ref{tab:attack-performance-4datasets}) are \underline{underlined}. The poisoning rates vary per dataset: Var.$^\dagger$: 2/0.3/0.3/0.3\%. Var.$^\ddagger$: 5/0.7/0.4/5\%.}
\label{tab:results_feature_resnet18}
\begin{tabular}{l c c c c c c c c c cccccccc}
\toprule
\multirow{2}{*}{Model} & \multirow{2}{*}{PR (\%)} 
& \multicolumn{2}{c}{C10} 
& \multicolumn{2}{c}{C100} 
& \multicolumn{2}{c}{Tiny} 
& \multicolumn{2}{c}{Imagenette} \\
\cmidrule(lr){3-4} \cmidrule(lr){5-6} \cmidrule(lr){7-8} \cmidrule(lr){9-10}
& & SS$\downarrow$ & DSWD$\downarrow$ &
SS$\downarrow$ & DSWD$\downarrow$ &
SS$\downarrow$ & DSWD$\downarrow$ & 
SS$\downarrow$ & DSWD$\downarrow$\\
\midrule
\multirow{2}{*}{BadNets} 
& 5   & 0.507 & 2.01 & 0.511 & 1.27 & 0.295 & 1.067 & 0.588 & 1.78\\
& 0.3 & 0.355 & 2.54 & 0.462 & 1.29 & 0.463 & 1.145 & 0.260 & 1.49\\
\cmidrule(lr){2-10}
\multirow{2}{*}{Blend} 
& 5   & 0.473 & 2.05 & 0.435 & 1.27 & 0.339 & 1.339 & 0.022 & 1.60\\
& 0.3 & 0.276 & 2.08 & 0.379 & 1.32 & 0.423 & 1.236 & 0.097 & 1.27\\
\cmidrule(lr){2-10}
\multirow{2}{*}{WaNet} 
& 5   & 0.382 & 1.83 & 0.492 & 1.21 & 0.342 & 1.301 & 0.557 & 1.73\\
& Var.$^\dagger$ & 0.378 & 1.62 & 0.217 & 0.877 & 0.303 & 0.926 & 0.192 & 0.474\\
\cmidrule(lr){2-10}
\multirow{2}{*}{Bpp} 
& 5   & 0.500 & 1.91 & 0.527 & 1.37 & 0.433 & 1.36 & 0.607 & 1.34\\
& Var.$^\dagger$ & 0.371 & 0.952 & 0.540 & 1.31 & \underline{0.347} & \best{\underline{0.049}} & \underline{0.227} & \best{\underline{0.040}}\\
\cmidrule(lr){2-10}
\multirow{2}{1cm}{Adap\\Patch} 
& 5   & \underline{0.254} & \best{\underline{0.765}} & 0.060 & 0.46 & \best{\underline{0.067}} & \underline{0.248} & \underline{0.082} & \underline{0.588}\\
& 0.3 & 0.0500 & 1.91 & \underline{0.058} & \best{\underline{0.224}} & \underline{0.245} & \underline{0.084} & \underline{-0.046} & \underline{0.106}\\
\cmidrule(lr){2-10}
\multirow{2}{1cm}{Adap\\Blend} 
& 5   & \underline{0.231} & \underline{1.15} & \best{0.038} & 0.693 & 0.213 & 0.717 & 0.0630 & 1.26\\
& 0.3 & 0.149 & 2.10 & 0.191 & 1.28 & 0.311 & 0.996 & \best{\underline{-0.066}} & \underline{0.965}\\
\cmidrule(lr){2-10}
\multirow{2}{*}{DFST} 
& 5   & \worst{0.519} & 2.26 & 0.509 & 1.34 & 0.485 & 1.707 & \worst{0.64} & 2.20\\
& 0.3 & 0.339 & 2.63 & \worst{0.730} & 1.93 & \worst{0.63} & 1.74 & 0.261 & 2.15\\
\cmidrule(lr){2-10}
\multirow{2}{*}{Narcissus} 
& Var.$^\ddagger$ & 0.491 & 2.23 & 0.074 & 0.945 & 0.314 & 1.977 & 0.043 & 1.34\\
& 0.3 & -0.007 & 1.90 & \underline{0.062} & \underline{0.852} & 0.302 & 1.703 & \underline{-0.016} & \underline{0.883}\\
\cmidrule(lr){2-10}
\multirow{2}{*}{Grond} 
& Var.$^\ddagger$ & 0.435 & 1.82 & \underline{0.233} & \underline{0.861} & 0.327 & 1.122 & 0.031 & 1.05\\
& 0.3 & \best{-0.101} & 1.65 & \underline{0.208} & \underline{0.786} & 0.355 & 1.071 & \underline{0.055} & \underline{0.286}\\
\cmidrule(lr){2-10}
DFBA & - & - & \worst{15.2} & - & \worst{18.0} & - & \worst{10.74} & - & \worst{4.64}\\
\bottomrule
\end{tabular}
\end{table}

\paragraph{Parameter-space footprints}
Table~\ref{tab:uclc-tac} reports the parameter-space footprint measurements for ResNet18 models.
Parameter-space metrics reveal a different picture. Grond generally produces small UCLC and TAC footprints, consistent with its explicit parameter-space objective. DFBA injects the backdoor into a single channel per layer. Therefore, it is conspicuous under UCLC and TAC, which respectively report the maximum value across layers and in the final convolutional layer. However, the TUP analysis in Section~\ref{sec:tup_case} shows that DFBA leaves a smaller footprint with a different choice of aggregation method and model layer. 

\renewcommand{\arraystretch}{0.7}
\begin{table}[t]
\scriptsize
\centering
\caption{Parameter-space footprint of backdoored ResNet18 models trained on our four datasets. The two metrics used to evaluate each model are the upper bound on the channel Lipschitz constant (UCLC) and trigger-activated change (TAC), both of which have the ideal value 0. For each metric and dataset, the \best{best} and \worst{worst} results are highlighted. Results that may be affected by the low ASR of the corresponding models (see Table~\ref{tab:attack-performance-4datasets}) are \underline{underlined}. *The poisoning rates vary per dataset.}
\label{tab:uclc-tac}
\setlength\tabcolsep{1pt}
\begin{tabular}{l c c c c c c c c c}
\toprule
\multirow{2}{*}{Model} & \multirow{2}{*}{PR (\%)} 
& \multicolumn{2}{c}{C10} 
& \multicolumn{2}{c}{C100} 
& \multicolumn{2}{c}{Tiny} 
& \multicolumn{2}{c}{Imagenette} \\
\cmidrule(lr){3-4} \cmidrule(lr){5-6} \cmidrule(lr){7-8} \cmidrule(lr){9-10}
& & UCLC$\downarrow$ & TAC$\downarrow$ & UCLC$\downarrow$ & TAC$\downarrow$ & UCLC$\downarrow$ & TAC$\downarrow$ & UCLC$\downarrow$ & TAC$\downarrow$\\
\midrule
Benign & - & 6.09 & - & 4.92 & - & 4.19 & - & 5.31 & - \\
\cmidrule(lr){2-10}
\multirow{2}{*}{BadNets} 
& 5   & 6.59 & 3.90 & 4.64 & 7.32 & 4.52 & 21.34 & 6.22 & 15.3 \\
& 0.3 & 7.99 & 8.03 & 5.86 & 9.17 & 4.57 & 38.89 & 5.39 & 18.4 \\
\cmidrule(lr){2-10}
\multirow{2}{*}{Blend} 
& 5   & 6.49 & 3.05 & 4.73 & 6.46 & 4.54 & 23.87 & 4.99 & 6.24 \\
& 0.3 & 5.76 & 4.30 & 4.92 & 9.69 & 4.85 & 25.31 & 5.37 & 6.04 \\
\cmidrule(lr){2-10}
\multirow{2}{*}{WaNet} 
& 5   & 6.33 & 10.7 & 6.52 & 20.9 & 4.27 & 31.08 & 7.05 & 64.0 \\
& 2/0.3/0.3/0.3$^*$ & 7.68 & 14.3 & 6.39 & 11.2 & 4.02 & 30.13 & 6.49 & 2.55 \\
\cmidrule(lr){2-10}
\multirow{2}{*}{Bpp} 
& 5   & 9.42 & 8.50 & 8.76 & 16.3 & 7.9 & 32.41 & 8.82 & 59.2 \\
& 2/0.3/0.3/0.3$^*$ & 7.38 & 7.48 & 7.21 & 10.0 & \best{\underline{3.95}} & \underline{10.63} & \underline{5.19} & \best{\underline{0.394}} \\
\cmidrule(lr){2-10}
\multirow{2}{1cm}{Adap\\Patch} 
& 5   & \underline{10.7} & \underline{6.44} & 5.35 & 6.83 & \underline{4.14} & \underline{14.88} & \underline{6.24} & \underline{11.5} \\
& 0.3 & 9.91 & 3.94 & \underline{6.06} & \best{\underline{1.45}} & \underline{5.57} & \best{\underline{10.44}} & \underline{7.62} & \underline{1.42} \\
\cmidrule(lr){2-10} 
\multirow{2}{1cm}{Adap\\Blend} 
& 5   & \underline{5.60} & \best{\underline{2.14}} & 6.32 & 6.62 & 4.41 & 21.18 & 5.08 & 5.61 \\
& 0.3 & 7.70 & 4.58 & 5.51 & 12.4 & 4.14 & 20.54 & \underline{5.67} & \underline{4.71} \\
\cmidrule(lr){2-10}
\multirow{2}{*}{DFST} 
& 5   & 6.16 & 6.41 & 4.28 & 11.9 & 4.35 & 26.05 & 6.29 & 18.9 \\
& 0.3 & 6.37 & 11.8 & 5.69 & 15.4 & 5.12 & 30.0 & 4.61 & 10.8 \\
\cmidrule(lr){2-10}
\multirow{2}{*}{Narcissus} 
& 5/0.7/0.4/5$^*$ & 7.37 & 6.23 & 6.19 & 6.25 & 4.28 & 35.86 & 5.41 & 5.59 \\
& 0.3 & 6.88 & 2.15 & \underline{5.44} & \underline{3.78} & 4.33 & 27.24 & \underline{6.42} & \underline{4.64} \\
\cmidrule(lr){2-10}
\multirow{2}{*}{Grond} 
& 5/0.7/0.4/5$^*$ & 3.60 & 2.43 & \best{\underline{3.78}} & \underline{6.00} & 4.33 & 21.89 & \best{3.14} & 4.20 \\
& 0.3 & \best{3.18} & 2.30 & \underline{4.10} & \underline{4.32} & 4.33 & 23.79 & \underline{3.73} & \underline{2.19} \\
\cmidrule(lr){2-10}
DFBA & - & \worst{17.8} & \worst{413} & \worst{17.7} & \worst{413} & \worst{21.17} & \worst{959.33} & \worst{18.7} & \worst{959} \\
\bottomrule
\end{tabular}
\end{table}

\paragraph{Takeaway}
The benchmark shows that small footprints are space-specific. BppAttack and WaNet often have small input-space footprints, but this does not make them uniformly quiet inside the model. Adaptive attacks can reduce feature-space separability, yet their effectiveness is less stable, and their other footprints remain visible. Grond tends to have small parameter-space footprints, consistent with its parameter-space objective, but this does not imply small input-space footprints. Thus, no attack family dominates across all three spaces.

\paragraph{Architecture stability} Table~\ref{tab:attack-performance-3arch} reports attack effectiveness across ResNet18, VGG16, and ViT-Small on CIFAR-10.
Input-space footprints are mostly determined by the trigger transformation rather than by the victim model, so we do not repeat input-space measurements for other architectures. 
We do provide the feature space metrics (SS and DSWD) for VGG16 and ViT-Small in Table~\ref{tab:feature-vgg-vit}; DFST is absent from these comparisons because it was not trained on VGG16 or ViT-Small, and DFBA has no ViT-Small variant.
Parameter space evaluation for VGG16 is provided in Table~\ref{tab:parameter-vgg}; VGG16 parameter-space data exists for only three attacks, too few for a meaningful rank correlation.
Parameter space evaluation for ViT is not offered because UCLC and TAC are designed based on the convolutional channel for CNN architectures, which is not compatible with the ViT architecture.
These architecture results should be read as a stability check rather than as a separate benchmark. The VGG16 results are broadly consistent with the ResNet18 analysis: feature-space rankings agree between the two convolutional networks (combined-feature Kendall $\tau{=}0.73$, Spearman $\rho{=}0.85$ over the eight attacks trained on all three architectures), Grond remains comparatively small in parameter space, and DFBA remains highly concentrated under UCLC and TAC. The ViT-Small feature-space results are less directly aligned with the CNN results (combined-feature $\tau{=}0.35$ against ResNet18), especially for attacks whose ASR drops substantially, such as WaNet and BppAttack. This reinforces the ASR caution used throughout the paper: a small feature-space footprint on a weak or failed backdoor should not be interpreted as strong stealthiness. Overall, the architecture check supports the tri-space conclusion while also showing that individual rankings can be architecture-dependent. The corresponding cross-architecture feature-space rankings are visualized in Appendix~\ref{app:architecture_ranking}, Figure~\ref{fig:architecture_feature_ranks}.

\begin{table}[htbp]
    \centering
    \footnotesize
    \setlength{\tabcolsep}{2.8pt}
    \caption{Performance of three architectures on CIFAR-10.}
    \label{tab:attack-performance-3arch}
    \begin{tabular}{l *{6}{c}}
    \toprule
    Attack & \multicolumn{2}{c}{ResNet18} & \multicolumn{2}{c}{VGG16} & \multicolumn{2}{c}{ViT-Small} \\
    \cmidrule(lr){2-3} \cmidrule(lr){4-5} \cmidrule(lr){6-7}
    ~ & BA & ASR & BA & ASR & BA & ASR \\
    \midrule
    Benign     & 94.7 & --   & 93.50 & --    & 80.62 & --    \\
    BadNets    & 94.5 & 100  & 93.45 & 100   & 80.94 & 100   \\
    Blend    & 94.3 & 99.8 & 93.43 & 99.66 & 81.10 & 98.96 \\
    WaNet      & 93.4 & 98.5 & 88.94 & 99.40 & 79.00 & 22.12 \\
    BppAttack        & 94.4 & 99.8 & 93.08 & 59.52 & 79.29 & 3.47  \\
    Adap-Patch     & 93.9 & 30.1 & 93.87 & 28.73 & 79.21 & 23.35 \\
    Adap-Blended     & 94.0 & 53.9 & 94.08 & 64.58 & 79.93 & 83.51 \\
    DFST       & 95.0 & 100  & --    & --    & --    & --    \\
    Narcissus    & 94.4 & 100  & 93.50 & 100   & 79.68 & 100   \\
    Grond      & 94.7 & 92.9 & 93.20 & 85.00 & 80.60 & 100   \\
    DFBA       & 94.3 & 100  & 93.20 & 100   & --    & --    \\
    \bottomrule
    \end{tabular}
\end{table}


\begin{table}[htbp]
    \centering
    \footnotesize
    \setlength{\tabcolsep}{5pt}
    \caption{Feature-space footprint measurements for backdoored VGG16 and ViT-Small models on CIFAR-10.}
    \label{tab:feature-vgg-vit}
    \begin{tabular}{l *{4}{c}}
    \toprule
    Attack & \multicolumn{2}{c}{VGG16} & \multicolumn{2}{c}{ViT-Small} \\
    \cmidrule(lr){2-3} \cmidrule(lr){4-5}
    ~ & SS & DSWD & SS & DSWD \\
    \midrule
    Badnet  & .515 & 2.784 & .263 & 2.719 \\
    Blended & .524 & 2.712 & .433 & 2.637 \\
    Wanet   & .398 & 1.978 & .195 & .290  \\
    Bpp     & .330 & 1.171 & .174 & .022  \\
    Adap-P  & .257 & .742  & .220 & .609  \\
    Adap-B  & .243 & 1.284 & .296 & 1.203 \\
    Narcis. & .459 & 3.360 & .467 & 3.431 \\
    Grond   & .317 & 2.340 & .375 & 2.114 \\
    DFBA    & --   & 3.900 & --   & --    \\
    \bottomrule
    \end{tabular}
\end{table}

\begin{table}[htbp]
    \centering
    \footnotesize
    \setlength{\tabcolsep}{2.5pt}
    \caption{Parameter-space footprint measurements for backdoored VGG16 models trained on three datasets.}
    \label{tab:parameter-vgg}
    \begin{tabular}{l c *{6}{c}}
    \toprule
    Model & PR & \multicolumn{2}{c}{CIFAR-10} & \multicolumn{2}{c}{CIFAR-100} & \multicolumn{2}{c}{Imagenette} \\
    \cmidrule(lr){3-4} \cmidrule(lr){5-6} \cmidrule(lr){7-8}
    ~ & ~ & UCLC & TAC & UCLC & TAC & UCLC & TAC \\
    \midrule
    Benign    & -- & 7.75 & -- & 4.63 & -- & 4.66 & -- \\
    Narcissus & 5/0.7/5$^*$ & 13.5 & 1.39 & 4.74 & 1.98 & 6.67 & 1.89 \\
    Grond     & 5/0.7/5$^*$ & 4.12 & .608 & 4.77 & 1.74 & 3.78 & 1.04 \\
    DFBA      & -- & 20.7 & 5.67 & 15.7 & 13.0 & 19.1 & 58.3 \\
    \bottomrule
    \end{tabular}
\end{table}

\section{Diagnostic Metrics and Failure Modes}
\label{sec:diagnostic_metrics}

In this section, we explore two recurring measurement blind spots in our benchmark. First, global feature-cluster metrics can be difficult to interpret when poisoned samples are class-fragmented or heavily imbalanced. Second, parameter-space metrics can emphasize channels that are high-gain but not trigger-relevant, or trigger-responsive but not downstream influential. We introduce CDBI and TUP as diagnostic additions for these cases. They are not intended to replace SS, DSWD, UCLC, or TAC; rather, they make specific footprint hypotheses easier to test. Table~\ref{tab:results_newmetric_resnet18} reports the full CDBI and TUP measurements for all ResNet18 configurations; as in the benchmark tables of Section~\ref{sec:benchmark_results}, both metrics are reported in their native orientations. These values underlie the diagnostic discussion in this section and the CDBI and TUP rankings in Table~\ref{tab:asr_sensitivity}.

\begin{table}[t]
\renewcommand{\arraystretch}{0.7}
\scriptsize
\centering
\caption{Full CDBI and TUP measurements of backdoored ResNet18 models trained on four datasets. Higher CDBI indicates stronger class-conditional feature overlap, i.e., a smaller measured feature-space footprint under this diagnostic (0 is the worst possible measurement); lower TUP indicates a smaller concentrated parameter-space footprint. For each metric and dataset, the \best{best} and \worst{worst} results are highlighted. Results that may be affected by the low ASR of the corresponding models (see Table~\ref{tab:attack-performance-4datasets}) are \underline{underlined}. CDBI is undefined for the data-free DFBA, which poisons no training samples. *The poisoning rates vary per dataset.}
\label{tab:results_newmetric_resnet18}
\setlength\tabcolsep{1pt}
\begin{tabular}{l c c c c c c c c c}
\toprule
\multirow{2}{*}{Model} & \multirow{2}{*}{PR (\%)}
& \multicolumn{2}{c}{C10}
& \multicolumn{2}{c}{C100}
& \multicolumn{2}{c}{Tiny}
& \multicolumn{2}{c}{Imagenette} \\
\cmidrule(lr){3-4} \cmidrule(lr){5-6} \cmidrule(lr){7-8} \cmidrule(lr){9-10}
& & CDBI$\uparrow$ & TUP$\downarrow$ & CDBI$\uparrow$ & TUP$\downarrow$ & CDBI$\uparrow$ & TUP$\downarrow$ & CDBI$\uparrow$ & TUP$\downarrow$\\
\midrule

    \multirow{2}{*}{BadNets} & 5 & 0.574 & 16.2 & 0.424 & 27.7 & 0.646 & 10.227 & \worst{0.418} & 18.4 \\

    & 0.3 & 0.627 & \worst{23.1} & 0.534 & 29.6 & 0.624 & 3.895 & 0.871 & \worst{19.3} \\
    \cmidrule(lr){2-10}

    \multirow{2}{*}{Blend} & 5 & 0.738 & 3.68 & 0.620 & 10.2 & 0.53 & 3.885 & 3.07 & 3.30 \\

    & 0.3 & 0.796 & 3.94 & 0.911 & 10.8 & 0.576 & 3.407 & 2.20 & 2.77 \\
    \cmidrule(lr){2-10}

    \multirow{2}{*}{WaNet} & 5 & 0.641 & 9.30 & 0.409 & 25.4 & 0.523 & 7.078 & 0.571 & 8.85 \\

    & 2/0.3/0.3/0.3$^*$ & 0.644 & 8.56 & 1.54 & 22.1 & 1.377 & 3.657 & 0.809 & 3.84 \\
    \cmidrule(lr){2-10}

    \multirow{2}{*}{Bpp} & 5 & 0.640 & 21.0 & 0.477 & 52.4 & 0.584 & \worst{34.275} & 0.518 & 13.9 \\

    & 2/0.3/0.3/0.3$^*$ & \worst{0.519} & 10.9 & 0.436 & \worst{56.1} & \underline{0.637} & \underline{3.578} & \underline{0.854} & \underline{1.56} \\
    \cmidrule(lr){2-10}

    \multirow{2}{1cm}{Adap\\Patch} & 5 & \underline{1.29} & \underline{8.38} & 1.19 & 13.4 & \best{\underline{1.841}} & \underline{3.522} & \underline{2.90} & \underline{5.72} \\

    & 0.3 & 3.42 & 10.2 & \underline{2.27} & \underline{9.77} & \underline{0.622} & \underline{3.495} & \underline{4.45} & \underline{2.35} \\
    \cmidrule(lr){2-10}

    \multirow{2}{1cm}{Adap\\Blend} & 5 & \underline{1.66} & \underline{3.34} & 1.24 & 9.93 & 0.916 & 3.209 & 3.00 & 3.29 \\

    & 0.3 & 5.19 & 3.63 & 1.53 & 11.2 & 0.497 & 3.451 & \underline{12.3} & \underline{2.45} \\
    \cmidrule(lr){2-10}

    \multirow{2}{*}{DFST} & 5 & 0.650 & 3.70 & 0.475 & 8.91 & 0.475 & \best{3.063} & 0.476 & 3.81 \\

    & 0.3 & 0.550 & 4.47 & \worst{0.260} & 11.1 & \worst{0.353} & 3.093 & 0.934 & 4.43 \\
    \cmidrule(lr){2-10}

    \multirow{2}{*}{Narcissus} & 5/0.7/0.4/5$^*$ & 0.767 & 2.15 & 3.47 & 8.75 & 1.107 & 3.192 & 4.27 & 2.64 \\

    & 0.3 & \best{5.92} & 2.13 & \best{\underline{4.34}} & \best{\underline{8.48}} & 1.434 & 3.25 & \underline{7.54} & \underline{2.13} \\
    \cmidrule(lr){2-10}

    \multirow{2}{*}{Grond} & 5/0.7/0.4/5$^*$ & 0.868 & 1.87 & \underline{1.95} & \underline{8.86} & 1.074 & 3.236 & \best{24.0} & 1.61 \\

    & 0.3 & 5.81 & \best{1.78} & \underline{1.53} & \underline{8.64} & 1.19 & 3.368 & \underline{12.4} & \best{\underline{1.05}} \\
    \cmidrule(lr){2-10}

    DFBA & -- & -- & 10.2 & -- & 10.6 & -- & 3.917 & -- & 13.5 \\
\bottomrule
\end{tabular}
\end{table}

\subsection{CDBI: Class-Conditional Feature Overlap}
\label{sec:cdbi_definition}

CDBI adapts the Davies--Bouldin index~\cite{davies1979dbi} to complement global cluster metrics by comparing poisoned source-class groups against the benign target-class cluster. Let $B_t$ be the embedded benign samples from target class $t$, and let $P_p$ be poisoned samples originating from source class $p$. 
For each source class, we compute $DBI_{t,p} = \frac{S(B_t) + S(P_p)}{M(B_t,P_p)}$, where
\[
S(A) = \frac{1}{|A|}\sum_{z \in A}\|z-\mu_A\|_2, \quad M(B_t,P_p)=\|\mu_{B_t}-\mu_{P_p}\|_2.
\]
We then average over source classes: $CDBI = \frac{1}{|\mathcal{P}|}\sum_{p\in \mathcal{P}} DBI_{t,p}$.
Higher CDBI indicates stronger class-conditional overlap and therefore a smaller feature-space footprint under this diagnostic. CDBI is computed on the same t-SNE embeddings of penultimate-layer features used for SS, with perplexity 30 (see Appendix~\ref{app:robust:cdbi}).

\subsection{When Does CDBI Differ from SS?}
\label{sec:cdbi_case}

CDBI and SS are related feature-space diagnostics, but they emphasize different geometric properties. SS treats clean and poisoned samples as two global clusters. This can penalize poisoned samples that fragment by source class, even when each source-class fragment overlaps with the benign target-class region. CDBI instead evaluates overlap class-conditionally.

WaNet illustrates the intended use case (e.g., in Appendix~\ref{sec:app:tsne_cifar10}, Figure~\ref{fig:cifar-tsne}, its poisoned samples form source-class subgroups, which lowers global cohesion and affects SS). CDBI instead asks whether each subgroup is close to the benign target-class cluster, leading to a different ranking. This disagreement is useful: it indicates that SS and CDBI are not interchangeable, and that the evaluator should inspect whether the apparent feature footprint is caused by clean/poisoned separation or by source-class fragmentation.

CDBI is not a standalone feature-stealthiness definition. Like the other feature-space metrics, it can be overly favorable for low-ASR attacks: triggered samples may remain close to clean features precisely because the backdoor was not successfully implanted, a limitation CDBI shares with SS and DSWD. Feature-space metrics should therefore always be interpreted together with ASR, and CDBI should be read alongside a global metric such as SS or DSWD, which capture complementary cluster geometry rather than correcting this confound.

\subsection{TUP: Joint Trigger Response and Downstream Influence}
\label{sec:tup_definition}

TAC measures trigger-induced activation change, while UCLC estimates downstream channel influence. Either signal alone can be misleading: a channel can be high-gain but unrelated to the trigger, or trigger-responsive but weakly connected to the output. TUP combines the two signals. Let $a^{(\ell)}_j(x)$ and $\hat{a}^{(\ell)}_j(x)$ denote the post-activation feature maps of channel $j$ at the analyzed final convolutional layer $\ell$ for a benign reference model trained on the same data and for the backdoored model, respectively. For the trigger-insertion operator $\mathcal{T}$,
\begin{align*}
TAC^{bkd}_j &= \mathbb{E}_{x}\,\big\|\hat{a}^{(\ell)}_j(\mathcal{T}(x)) - \hat{a}^{(\ell)}_j(x)\big\|_{2}, \\
TAC^{ben}_j &= \mathbb{E}_{x}\,\big\|a^{(\ell)}_j(\mathcal{T}(x)) - a^{(\ell)}_j(x)\big\|_{2},
\end{align*}
both computed over the same image set and normalization pipeline. For channel $j$, let
\[
R_j = \frac{TAC^{bkd}_j}{TAC^{ben}_j},
\qquad
TUP_j = R_j \cdot UCLC_j.
\]
The ratio $R_j$ cancels the trigger-magnitude component shared by the benign and backdoored models, so that large input perturbations alone do not inflate the parameter-space verdict. We report the mean or distribution of $TUP_j$ over channels in the analyzed final convolution layer. Lower values indicate a smaller concentrated parameter-space footprint under this diagnostic. TUP should therefore be interpreted as a diagnostic for concentrated trigger-sensitive high-impact channels, not as a complete parameter-space stealthiness definition.

\subsection{When Does TUP Differ from UCLC and TAC?}
\label{sec:tup_case}

UCLC and TAC measure different aspects of a parameter-space footprint. UCLC estimates downstream influence, but a high-UCLC channel need not be trigger-relevant. TAC measures trigger-induced activation change, but a high-TAC channel need not be influential for the final prediction. TUP tests the joint condition: whether a channel is both trigger-responsive and downstream influential.
Figure~\ref{fig:tup_scatter} visualizes this joint condition by plotting channels according to trigger-response ratio and downstream influence: high-TUP channels are those that are large on both axes, not merely large under one metric alone.

This distinction matters in our benchmark. Grond generally has small UCLC, TAC, and TUP values, consistent with its explicit parameter-space objective (Table~\ref{tab:results_newmetric_resnet18}). DFBA, by contrast, is conspicuous under UCLC and TAC, but its TUP values ($10.2$, $10.6$, $3.9$, and $13.5$ on the four datasets) are exceeded by several poisoning attacks, because TUP is computed over the final convolutional layer and averaged over channels. If the most trigger-relevant or most influential channel lies outside this layer, or if the backdoor is concentrated in a single channel while the reported statistic averages over many channels, TUP can understate the footprint.

\definecolor{highTUP}{HTML}{fde725}
\definecolor{midTUP}{HTML}{a0f43c}
\definecolor{lowTUP}{HTML}{440154}

\begin{figure}
    \centering
    \includegraphics[width=\linewidth]{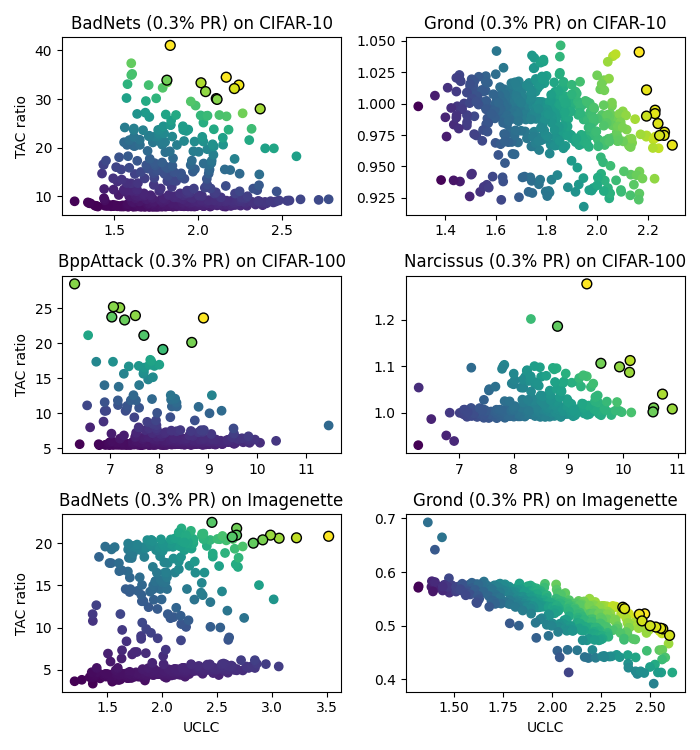}
    \caption{TAC-UCLC Product (TUP) measurements for each channel in the final layer of backdoored models. Channels are colored from \textcolor{lowTUP}{lowest} \textcolor{midTUP}{to} \textcolor{highTUP}{highest} TUP, and the top-10 channels are outlined.}
    \label{fig:tup_scatter}
\end{figure}

TUP is therefore not a replacement for UCLC or TAC. It is a diagnostic for a narrower hypothesis: whether the analyzed layer contains channels that are simultaneously trigger-responsive and downstream influential. This is useful for distinguishing high-gain but trigger-irrelevant channels from channels more plausibly involved in the backdoor mechanism, but it also has a clear limitation shared by the TAC metric: layer choice matters.

\begin{table*}[t]
\centering
\caption{Representative diagnostic disagreements. CDBI and TUP are not replacements for existing metrics; they identify specific cases where global feature or parameter metrics can miss important structure.}
\label{tab:diagnostic_examples}
\begin{tabular}{lll}
\toprule
Case & Observation & Interpretation \\
\midrule
WaNet vs. SS & Poisoned samples split by source class & CDBI checks class-conditional overlap \\
DFBA vs. TUP & High UCLC/TAC but layer-dependent TUP & TUP depends on analyzed layer \\
Low-ASR attacks & Small footprints can coincide with failed backdoors & Interpret with ASR \\
\bottomrule
\end{tabular}
\end{table*}

\paragraph{Summary}
Table~\ref{tab:diagnostic_examples} summarizes the representative disagreements discussed in this section. These diagnostic disagreements are not edge cases; they illustrate the kinds of failures that appear repeatedly in multi-space measurement. WaNet shows why global feature clustering can be difficult to interpret when poisoned samples fragment by source class. DFBA shows why parameter-space metrics should distinguish trigger response from downstream influence. Low-ASR configurations show why no footprint metric should be interpreted without attack effectiveness. These examples motivate treating CDBI and TUP as diagnostic additions: they help localize why an existing metric gives a particular verdict, but they do not replace the broader metric suite.
\section{Are Stealthiness Metrics Interchangeable?}
\label{sec:metric_disagreement}

A prerequisite for any stealthiness benchmark is to know whether its metrics are interchangeable estimates of one underlying property or whether they capture distinct, non-substitutable footprints. We test this directly. We orient all footprint metrics so that a larger value always denotes a larger footprint, inverting the metrics for which higher is conventionally better, such as PSNR, SSIM, pHash, and CDBI. Treating each backdoored ResNet18 configuration (dataset $\times$ attack $\times$ poisoning rate) as one observation, we compute pairwise Spearman rank correlations. We use Spearman correlation because we care about agreement on the ordering of attacks, not the linear scale of any metric.

\begin{figure*}[t]
    \centering
    \includegraphics[width=0.92\linewidth]{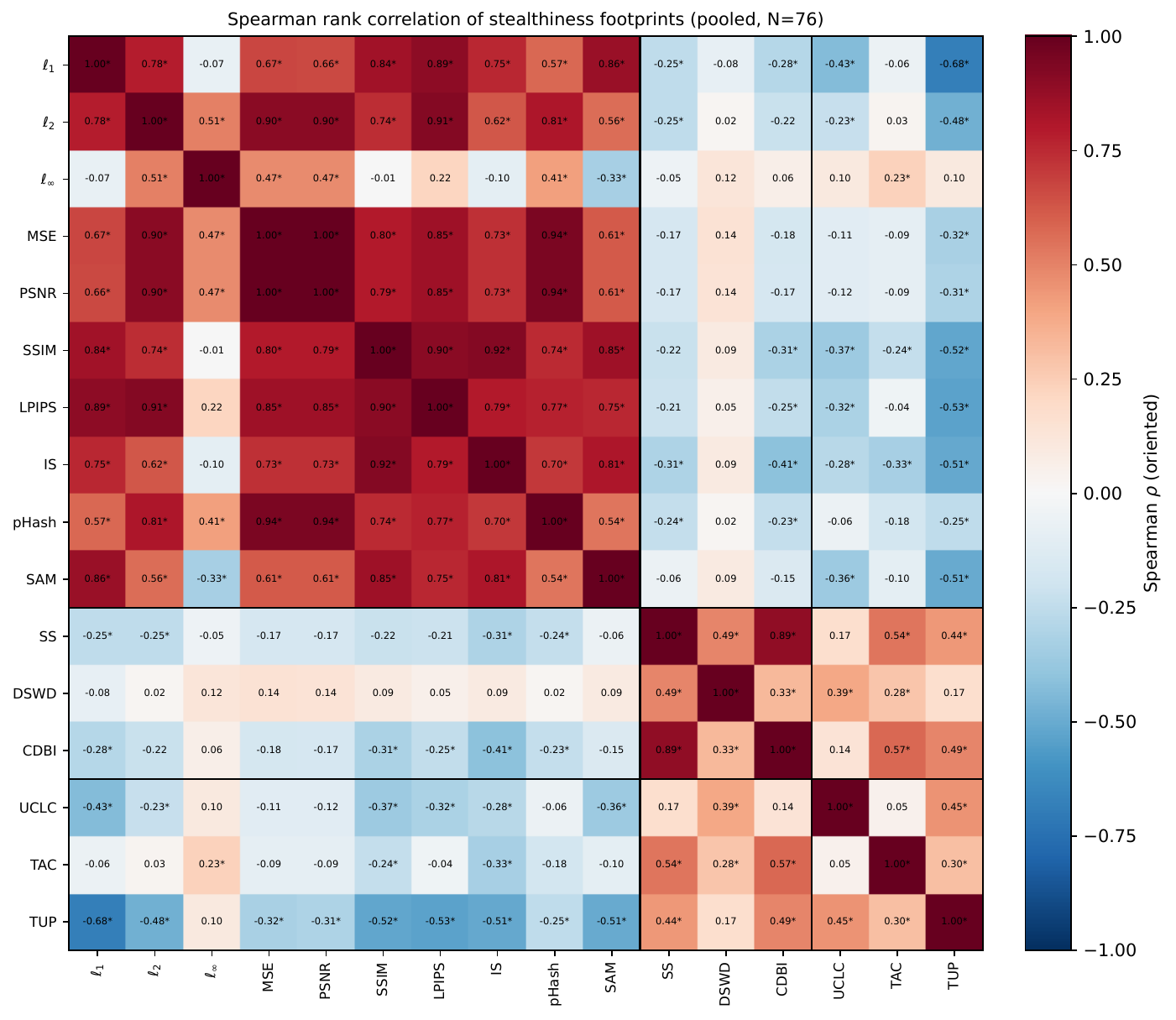}
    \caption{Spearman rank correlation among footprint metrics after
orienting every metric so that larger values indicate larger measured
footprints. Each backdoored ResNet18 configuration
(dataset $\times$ attack $\times$ poisoning rate, $N=76$) is one
observation; $^\ast$ marks $p<0.05$. Black lines separate input-,
feature-, and parameter-space metrics. Metrics cluster within spaces
but largely disagree across them: the input block is highly redundant,
whereas input--feature and input--parameter correlations are near zero
or negative. This supports tri-space reporting rather than a single
stealthiness scalar.}
    \label{fig:metric_correlation}
\end{figure*}

Figure~\ref{fig:metric_correlation} visualizes the full pooled correlation matrix. Table~\ref{tab:metric_correlation_blocks} summarizes average within- and across-space agreement. We observe that the metrics are not interchangeable, and agreement is structured by observation space rather than by a single latent stealthiness axis. Agreement is strongest within the input space, where several metrics are near-duplicates: MSE and PSNR are rank-identical by construction, and SSIM and LPIPS often agree strongly. The main exception is $\ell_\infty$, which can be weakly correlated with or opposed to other input metrics because patch triggers perturb a few pixels maximally while warping or blending triggers spread smaller changes over the whole image. Agreement weakens within the feature space and is weakest within the parameter space, where UCLC and TAC can rank attacks differently because they measure downstream influence and trigger-induced activation change, respectively.

Across spaces, the divergence is starker. Input--feature and input--parameter correlations are near zero or negative. This quantifies the central trade-off in our title: attacks that are quiet in the input space can be loud in the parameter space, while attacks with smaller parameter footprints can require larger input perturbations. The only non-trivial cross-space agreement appears between some feature- and parameter-space metrics, but this agreement weakens when low-ASR configurations are filtered out (restricting to configurations with ASR $\geq 60\%$ reduces the pooled feature--parameter correlation from $+0.35$ to $+0.29$), suggesting that part of the apparent agreement is driven by attack effectiveness rather than by a shared stealthiness signal; Section~\ref{sec:asr_effectiveness} isolates this effect. Figure~\ref{fig:metric_correlation} and Table~\ref{tab:metric_correlation_blocks} themselves use all $N{=}76$ configurations without ASR filtering.

Two clarifications scope these results. First, SS and CDBI are strongly rank-correlated overall ($\rho\approx0.89$ pooled, $0.78$ within-dataset), confirming that CDBI is a feature-space metric in the same family as SS rather than an unrelated quantity; the regimes in which the two diverge, and in which CDBI is the more informative diagnostic, are examined in Section~\ref{sec:cdbi_case}. Second, pooling across datasets with very different activation scales deflates parameter-space agreement (within-dataset $+0.50$ versus pooled $+0.27$ in Table~\ref{tab:metric_correlation_blocks}), so within-dataset correlations are the more conservative summary of parameter-space agreement.

Taken together, these results establish the central premise of our study: a small footprint under one metric, or in one space, does not imply a small footprint elsewhere. Stealthiness is therefore a multi-view measurement problem, and a claim of stealthiness must state which footprint it refers to rather than treating stealthiness as a single scalar.

\begin{table}[t]
\centering
\caption{Average Spearman rank correlation among footprint metrics
within and across spaces. Within-dataset correlations are computed
separately for each dataset and then averaged to reduce cross-dataset
scale effects; pooled correlations treat all configurations as one
sample. Agreement is high within input space, weaker within feature
and parameter spaces, and low or negative across spaces.}
\label{tab:metric_correlation_blocks}
\begin{tabular}{lcc}
\toprule
Metric group & Within-dataset $\rho$ & Pooled $\rho$ \\
\midrule
Input--Input & +0.69 & +0.66 \\
Feature--Feature & +0.61 & +0.57 \\
Parameter--Parameter & +0.50 & +0.27 \\
Input--Feature & $-0.09$ & $-0.11$ \\
Input--Parameter & $-0.28$ & $-0.24$ \\
Feature--Parameter & +0.42 & +0.35 \\
\bottomrule
\end{tabular}
\end{table}

The negative input--parameter correlations are especially important for interpreting stealthiness claims. They indicate that reducing visible or perceptual trigger footprints does not necessarily reduce internal concentration; in some cases, the two move in opposite directions. Conversely, the moderate feature--parameter association does not justify replacing one space with the other. A feature-space metric can detect representation separability, while a parameter-space metric can detect whether the backdoor relies on a small set of influential channels. These are different observations of the same model, and disagreement between them is informative rather than anomalous.

This also explains why single-space claims can be misleading. An input-stealthy attack can evade visual inspection while still inducing feature clusters or parameter concentration. A feature-stealthy attack can reduce clean/poisoned separability while still failing at low poisoning rates. A parameter-quiet attack can reduce channel-level evidence while still relying on an input perturbation that is measurable. The benchmark therefore treats metrics as complementary evidence, not as interchangeable proxies.
\section{Attack Effectiveness and Defense Checks}
\label{sec:robustness_external_checks}

\subsection{Footprints versus Attack Effectiveness}
\label{sec:asr_effectiveness}

Low-ASR configurations create an interpretation challenge: a small footprint may indicate either a quiet attack or a backdoor that was never successfully implanted. To test whether our conclusions depend on such cases, we recompute the per-space attack rankings after excluding configurations with ASR below 60\% and 80\%; this removes 15 and 25 of the 76 configurations, respectively. Rankings use within-dataset percentile-footprint aggregation averaged over each attack's surviving configurations.
Table~\ref{tab:asr_sensitivity} summarizes the ASR-filtering sensitivity analysis; the main conclusion is unchanged under both ASR filters.
Table~\ref{tab:asr_sensitivity} also shows why we avoid treating low-footprint low-ASR configurations as evidence of successful stealthiness. The per-space winners remain stable after filtering, but the strict ASR threshold reduces the number of usable configurations for some attacks. This means that the main qualitative conclusion is robust, while fine-grained rankings under the strictest filter should be interpreted cautiously. The important point is not that the exact ordering of every attack is invariant, but that the tri-space pattern remains: input-, feature-, and parameter-space footprints identify different attacks, and no single attack has the smallest measured footprint across all three spaces.


\begin{table}[htbp]
\centering
\footnotesize
\caption{Sensitivity of the main conclusions to ASR filtering. Per-space rankings use
within-dataset percentile footprints averaged over each attack's surviving configurations
(lower values indicate smaller measured footprints). The per-space winners and the tri-space conclusion are unchanged when
low-ASR (failed) attacks are excluded.}
\label{tab:asr_sensitivity}
\setlength{\tabcolsep}{4pt} 
\begin{tabularx}{\linewidth}{@{} >{\raggedright\arraybackslash}X c c c @{}}
\toprule
\textbf{Conclusion} & \textbf{All configs} & \textbf{ASR$\ge$60\%} & \textbf{ASR$\ge$80\%} \\
\midrule
Smallest input-space footprint         & BppAttack     & BppAttack     & BppAttack \\
Smallest feature-space footprint       & Adap-Patch    & Adap-Patch    & Adap-Patch \\
Smallest parameter-space footprint     & Grond         & Grond         & Grond \\
Same attack best in all three spaces?  & No            & No            & No \\
Most balanced attack (min worst space) & Grond ($0.42$) & Grond ($0.45$) & Grond ($0.36$) \\
CDBI top-ranked attack                 & Narcissus     & Narcissus     & Narcissus \\
TUP top-ranked attack                  & Grond         & Narcissus     & Grond \\
\bottomrule
\end{tabularx}
\end{table}

The central conclusions are stable. The attack with the smallest measured footprint in each space is unchanged across all three filters: BppAttack in input space, Adap-Patch in feature space, and Grond in parameter space. No attack has the smallest measured footprint across all three spaces under any filter. Even the most balanced attack, Grond, remains near the 42nd footprint percentile in its worst space, so no attack occupies the low-footprint tail of all three spaces simultaneously. Rank stability is high within each space: Kendall's $\tau$ between the unfiltered and filtered orderings is $+0.96/+0.91$ for the input and feature spaces and $+0.94/+0.76$ for the parameter space at the $60\%/80\%$ thresholds.

Two points scope these results. First, the strictest filter is data-limited for attacks that rarely reach high ASR. Grond is well supported at ASR $\geq 60\%$ (five surviving configurations), whereas Adap-Patch retains only two, and at ASR $\geq 80\%$ the feature- and parameter-space winners survive on a single high-ASR configuration each. Their rankings should therefore be interpreted alongside the number of surviving configurations. However, even these few successful configurations retain the smallest measured footprints in their respective spaces, so the per-space winners are not produced solely by failed runs.

Second, composite cross-space rankings are less stable ($\tau = +0.73/+0.56$) than individual-space rankings. This is expected rather than concerning, because the three spaces capture different footprints. This is precisely why we report per-space measurements rather than a single composite ranking. The diagnostic metrics behave consistently as well: CDBI identifies Narcissus as its top-ranked attack under all filters, and TUP identifies Grond, with Narcissus marginally overtaking only at the ASR $\geq 60\%$ threshold (Table~\ref{tab:asr_sensitivity}); because both attacks retain successful high-ASR configurations, these rankings are not artifacts of failed backdoors.

Overall, excluding failed attacks does not change the tri-space conclusion: no attack leaves the smallest measured footprint across all three spaces (see also Appendix~\ref{app:robust}).

\subsection{Defense Outcomes as External Checks}
\label{sec:defense_external_checks}

Defense behavior provides an important external check on footprint measurements: if a measured footprint is practically exploitable, defenses that target the corresponding signal should sometimes succeed. We therefore analyze associations between footprint metrics and defense outcomes. We treat this analysis as auxiliary rather than definitional. A defense that succeeds identifies one exploitable footprint, while a defense that fails does not imply that no footprint exists.

This distinction is necessary because defenses differ in access assumptions, thresholds, outputs, and targeted mechanisms. Some defenses reconstruct input triggers, some detect poisoned samples in feature space, and others prune or repair model components. Their outcomes are also confounded by attack effectiveness: an attack with low initial ASR may appear easy to mitigate for reasons unrelated to stealthiness. For this reason, we interpret defense--footprint associations as evidence about whether particular footprints are operationally useful, not as a universal stealthiness score.

The results support the tri-space view. Defense outcomes are defense- and attack-dependent: no defense provides a complete proxy for stealthiness, and associations are strongest when the defense mechanism aligns with the measured footprint. Input-oriented defenses are most informative about input-space footprints, feature-space detectors about representation-level separability, and pruning- or activation-based mitigations about parameter-space concentration. This external-check analysis evaluates the attacks against ten defenses on two benchmark datasets, CIFAR-10 and Imagenette. Table~\ref{tab:defense_summary} summarizes the defense categories; full per-attack defense outcomes are reported in the supplementary material, and the residual association analysis follows in Section~\ref{sec:defense_association}.

\begin{table*}[t]
\centering
\scriptsize
\caption{Summary of supplementary defense outcomes. Mitigation defenses are summarized by post-defense ASR, where lower is better for the defense. Detection defenses are summarized by F1, where higher is better for the defense. Full per-attack results are reported in
the supplementary material.}
\label{tab:defense_summary}
\begin{tabular}{llll}
\toprule
Defense type & Defenses & Output & Main use \\
\midrule
Input-trigger reconstruction & Neural Cleanse~\cite{wang2019nc}, TABOR~\cite{guo2019tabor}, BAN~\cite{xu2024ban} & Reversed trigger norm / anomaly & Input-footprint check \\
Feature/sample detection & SPECTRE~\cite{hayase2021spectre}, Spectral Signatures~\cite{tran2018spectral}, STRIP~\cite{gao2020strip}, BTI-DBF~\cite{xu2024towards} & F1 or detected samples & Feature/input-footprint check \\
Model repair/mitigation & CLP~\cite{zheng2022clp}, FeatureRE~\cite{wang2022featurere}, I-BAU~\cite{zeng2022ibau} & Post-defense ASR & Parameter/behavioral check \\
\bottomrule
\end{tabular}
\end{table*}

\subsection{Defense--Footprint Association Analysis}
\label{sec:defense_association}

The defense summary in Section~\ref{sec:defense_external_checks} establishes why defense outcomes are auxiliary rather than definitional. We now ask a narrower empirical question: are measured footprints associated with defense outcomes after controlling for obvious confounders? This analysis is not used to define stealthiness. Its purpose is to test whether some footprints correspond to operational signals exploited by existing defenses.

We align all quantities so that larger metric values indicate smaller measured footprints and larger defense-outcome values indicate stronger defense evasion. For mitigation defenses, larger post-defense ASR means stronger evasion. For detection defenses, such as F1-based detectors, we invert the sign so that larger values correspond to weaker detection. We then regress both metric values and defense outcomes on dataset identity and pre-defense ASR, and compute Spearman correlations between the residuals.

\begin{figure*}[!htbp]
\centering
\includegraphics[width = 0.9\linewidth]{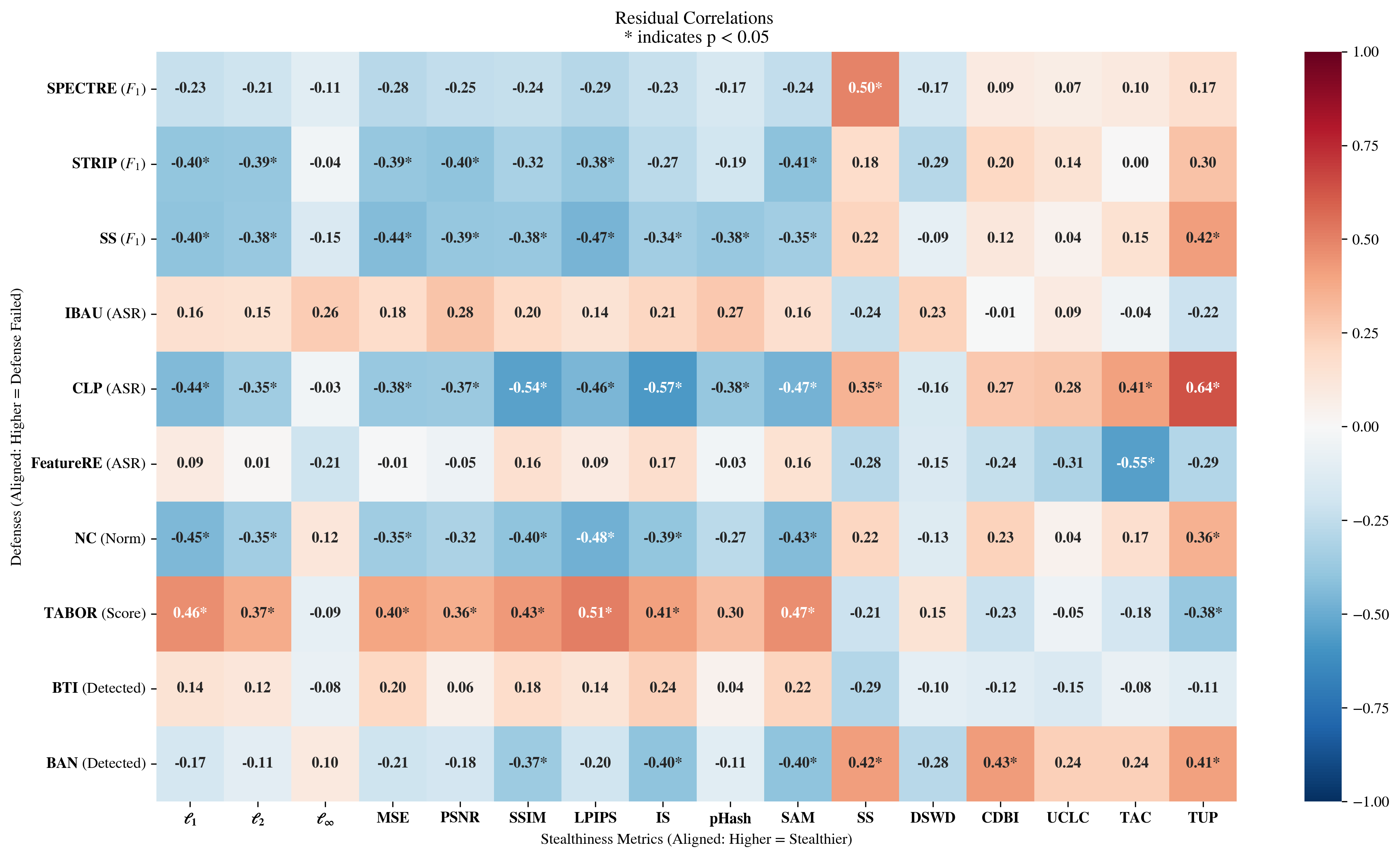}
\caption{Residual association between footprint metrics and defense outcomes after controlling for dataset identity and pre-defense ASR. Footprint metrics are oriented so that larger values indicate smaller measured footprints, while defense outcomes are oriented so that larger values indicate stronger defense evasion. Positive values, therefore, indicate that a smaller measured footprint is associated with stronger evasion of the corresponding defense. ``SS (F1)'' on the defense axis denotes the Spectral Signatures detector, distinct from the Silhouette Score (SS) on the metric axis. The associations are defense-specific rather than universal, supporting the use of defense outcomes as external checks rather than as definitions of stealthiness.}
\label{fig:residual_correlations}
\end{figure*}

Figure~\ref{fig:residual_correlations} shows that defense associations are structured by defense mechanism. Input-oriented defenses are most associated with input-space metrics, feature/sample detectors with feature-space metrics, and pruning or repair methods with parameter-space metrics. At the same time, no metric is uniformly predictive across all defenses. This supports the paper's main conclusion: defenses can exploit particular footprints, but no defense outcome provides a universal proxy for stealthiness.

A positive residual association means that a smaller measured footprint is associated with stronger evasion of a particular defense. A weak or negative association does not invalidate the metric; it can indicate that the defense targets another footprint, that thresholds are poorly matched, or that the relationship is attack-family-specific.
Table~\ref{tab:defense_summary} summarizes the defense categories, and the supplementary material reports the full defense outcomes used in the residual association analysis.

\section{Artifact, Recommendations, and Limitations}
\label{sec:guidance_limitations}

\subsection{Benchmark Artifact}
\label{sec:artifact}

To make the measurement protocol reusable, we organize the artifact as an extensible benchmark rather than a collection of one-off scripts. The benchmark separates attacks, models, datasets, and metrics through a small set of wrappers. Each attack implementation exports a backdoored model and poisoned train/test data, which are normalized into a common record containing the model, clean and poisoned datasets, poisoned-sample indicators, original labels, target labels, and attack metadata. The artifact also includes the backdoored ResNet18 models and the poisoned data for CIFAR-10, CIFAR-100, and Imagenette.

Metrics consume standardized artifacts rather than attack-specific internals. Input-space metrics operate on paired clean and poisoned image batches; feature-space metrics consume predictions and penultimate-layer representations; and parameter-space metrics consume model parameters and, when needed, saved activation differences. This design allows new attacks to be added by providing a wrapper, and new metrics to be added without reimplementing attacks.

Table~\ref{tab:artifact_components} summarizes the interfaces and stored outputs that make the benchmark auditable and extensible.

\begin{table}[htbp]
\centering
\footnotesize
\caption{Benchmark artifact components.}
\label{tab:artifact_components}
\begin{tabularx}{\linewidth}{@{} l >{\raggedright\arraybackslash}X @{}}
\toprule
\textbf{Component} & \textbf{Purpose} \\
\midrule
Attack wrapper & Normalizes trigger generation, poisoned data, and attack metadata \\
Model wrapper & Standardizes prediction, feature extraction, and parameter access \\
Metric API & Computes input-, feature-, and parameter-space footprint values \\
Run schema & Stores BA, ASR, metrics, configurations, and reproducibility metadata \\
Aggregation scripts & Produce benchmark tables and derived analyses from saved runs \\
Extension hooks & Allow new attacks or metrics to be added under the same protocol \\
\bottomrule
\end{tabularx}
\end{table}

The artifact records BA, ASR, dataset, architecture, poison rate, target class, random seed, attack configuration, and all measured footprints. It also stores intermediate artifacts such as extracted features, predictions, activation changes, and poisoned-sample indices. This supports two forms of reuse: auditing each aggregate result back to per-configuration measurements, and inserting new attacks or metrics into the same multi-space protocol. Our code, data, and supplementary material are publicly available at \url{https://doi.org/10.5281/zenodo.22757052}.

\subsection{Recommendations}
Our results suggest several practical guidelines for backdoor stealthiness evaluation. First, papers should report attack effectiveness and stealthiness evidence together. A low-footprint configuration with low ASR should not be presented as a successful stealthy attack without qualification. Second, stealthiness claims should specify the observation space: input, feature, parameter, or defense behavior. Third, at least one metric from each space should be reported when a paper claims broad stealthiness. Fourth, metric disagreement should be expected rather than treated as an error. Disagreement often identifies where an attack leaves a footprint.

\paragraph{Suggested reporting template}
Concretely, each attack evaluation should state the threat model and poisoning rate alongside benign accuracy and attack success rate, and then report the per-space evidence as follows. Input-space metrics should include both pixel-level and perceptual measurements when possible. Feature-space metrics should include a global separability metric and, when class fragmentation is plausible, a class-conditional diagnostic such as CDBI. Parameter-space metrics should include both trigger response and downstream influence when white-box access is available.

\subsection{Limitations}
The same measurement framing that makes the benchmark useful also bounds its interpretation: our results characterize observable footprints under a controlled protocol, not an absolute or deployment-independent notion of stealthiness. First, this benchmark is not an absolute attack leaderboard. The evaluated attacks differ in threat model, poisoning constraints, trigger construction, and training access. We standardize the evaluation environment where possible, but the results should be interpreted as controlled footprint measurements rather than definitive attack rankings.

Second, the metric suite is not exhaustive. Input-space metrics are imperfect proxies for human perception, feature-space metrics depend on the embedding and sample distribution, and parameter-space metrics depend on the architecture and layer choice. CDBI and TUP address specific failure modes, but they do not define stealthiness on their own.

Third, parameter-space measurement is currently strongest for convolutional architectures. Extending comparable measurements to transformers requires architecture-specific definitions of internal components, such as attention heads, MLP neurons, or transformer blocks. We therefore interpret the ViT results mainly through input- and feature-space footprints.

Fourth, white-box footprint metrics assume access to triggers, poisoned data, model internals, or activation traces. This is appropriate for benchmarking, auditing, and attack development, but it differs from the operational setting of a defender who may not know the trigger or poisoning process. Defense outcomes provide useful external checks, but they should not be treated as the definition of stealthiness.

Finally, any fixed benchmark can be gamed if attackers optimize directly against its metrics. For this reason, the main recommendation is not to standardize on one scalar score, but to report a diverse set of footprints and to make the measurement protocol extensible.
\section{Related Work}
\label{sec:related_work}

Early attacks such as BadNets~\cite{gu2019badnets} use visible patch triggers, making poisoned images easy to spot by inspection or by inference-time input screening (e.g., STRIP~\cite{gao2020strip}). STRIP identifies backdoors by superimposing input images with clean samples; consistently low entropy in predictions across blended inputs indicates a trigger's presence. Subsequent attacks pursued imperceptibility. WaNet~\cite{nguyen2021wanet}, Lira~\cite{doan2021lira}, BppAttack~\cite{wang2022bppattack}, among others, design invisible triggers by image processing techniques (image warping, quantization) or train a generator to find the optimal trigger, yielding small per-pixel changes but strong attack effectiveness.
These works established that input-space stealthiness is achievable and input-only defenses will miss sufficiently subtle triggers.

Although poisoned images may appear visually indistinguishable from clean ones, they can often be separated in the model’s latent space. Activation Clustering (AC)~\cite{chen2018ac} exploits this property by embedding activations, typically after dimensionality reduction, and identifying the presence of two clusters within a class, which indicates a target-class backdoor. This vulnerability has motivated the development of attacks that explicitly reduce latent separability. For instance, Qi et al. introduced Adap-Patch and Adap-Blend~\cite{qi2023adap}, which weaken the association between triggers and target classes such that triggered features more closely overlap with benign features, thereby evading AC in multiple settings. Similarly, WaNet and the adaptive attacks highlight an inherent trade-off: achieving stealthiness in the input space does not necessarily imply stealthiness in the feature space, and vice versa. Complementary research has investigated feature-space stealthiness under relaxed threat models, where tailored loss functions are designed to minimize the distance between benign and backdoored features~\cite{doan2021dswd,bypass2020tan}.

Parameter-space defenses ask whether a small subset of channels carries the backdoor. Channel Lipschitzness-based Pruning (CLP)~\cite{zheng2022clp} estimates a per-channel upper bound on sensitivity (UCLC) and prunes the most sensitive channels without using any data; many input/feature-stealthy attacks still fail under CLP. Xu et al.~\cite{xu2025grond} argued that contemporary attacks are rarely evaluated against such defenses and introduced  {Grond}, trained with Adversarial Backdoor Injection (ABI), which penalizes or replaces overly sensitive channels during training to diffuse the trigger’s influence. Grond aims to couple a small parameter-space footprint with imperceptible triggers, and its authors report resistance to several feature-space detectors. This motivates evaluating it across all three spaces; resistance to selected feature-space defenses does not by itself establish a uniformly small feature-space footprint.

Finally, broader studies also exist; for instance, Liu et al. evaluated multiple attacks with a large set of perceptual metrics~\cite{liu2022stealthiness}, but their focus is primarily on the input space. BackdoorBench~\cite{baoyuan2022bb} standardizes attack and defense implementations, with primary emphasis on attack effectiveness and defense evaluation; it includes some stealthiness metrics, yet most implemented attacks remain input-centric. Our benchmark is complementary: it standardizes the measurement of observable footprints across input, feature, and parameter spaces, preserves the intermediate artifacts required to audit those measurements, and directly analyzes whether the resulting metrics are interchangeable. The contribution is therefore not another attack repository, but a measurement layer that can be applied across heterogeneous attack implementations and extended with new footprint metrics.

\section{Conclusion}
\label{sec:conclusion}

This paper presented a tri-space measurement study of backdoor stealthiness. We distinguished latent stealthiness from observable footprints in input, feature, and parameter space, and evaluated representative backdoor attacks across datasets, architectures, and metrics. The central finding is that stealthiness is not monolithic: attacks that leave small footprints in one space often leave larger footprints in another, and existing metrics are not interchangeable.

We introduced CDBI and TUP as diagnostic additions for specific blind spots in feature-space and parameter-space measurement. CDBI helps interpret class-conditional feature overlap, while TUP identifies channels that are both trigger-responsive and downstream influential. These diagnostics augment rather than replace the broader metric suite.

Overall, our results argue against single-metric, defense-only, or aggregate-score claims of backdoor stealthiness. Future evaluations should report attack effectiveness together with multi-space footprint measurements, state which footprint is being claimed, and avoid collapsing stealthiness into a single universal score.




\section*{Acknowledgments}
This work has been partially funded by the European Union under the Horizon Europe research and innovation programme — SHASAI project (Secure Hardware and Software for AI Systems), Grant Agreement No. 101225866.

\bibliographystyle{IEEEtran}
\bibliography{bibliography}

\appendices

\section{t-SNE Embeddings for ResNet18 on CIFAR-10}
\label{sec:app:tsne_cifar10}

The t-SNE embeddings for ResNet18 on CIFAR-10 are shown in \autoref{fig:cifar-tsne}.

\newcommand{\tsnefigurecifar}[4][0.003]{
    \begin{subfigure}[b]{0.32\textwidth}
        \centering
        \caption{#4}
        \vspace{-0.15cm} 
        \includegraphics[width=0.48\linewidth]{tsne/#2/#3_p0.05/default.png}\hfill
        \includegraphics[width=0.48\linewidth]{tsne/#2/#3_p#1/default.png}
    \end{subfigure}
}

\captionsetup[subfigure]{labelformat=empty, skip=2pt} 
\begin{figure*}[htbp]
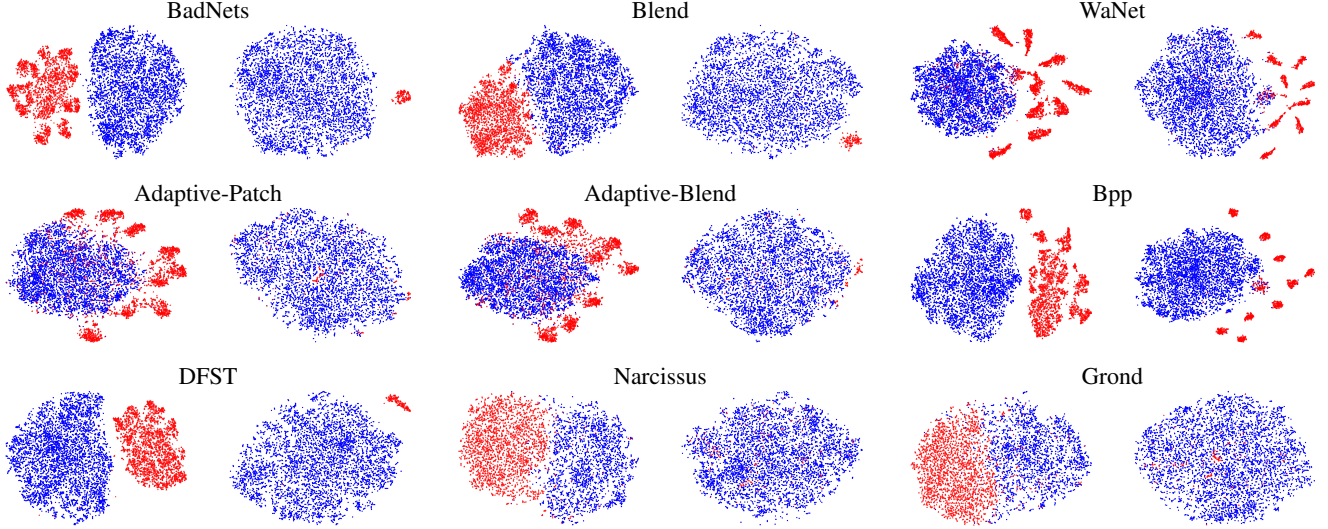

    \centering
    \tsnefigurecifar{resnet18_cifar10}{badnet}{BadNets}\hfill
    \tsnefigurecifar{resnet18_cifar10}{blended}{Blend}\hfill
    \tsnefigurecifar[0.02]{resnet18_cifar10}{wanet}{WaNet}
    
    \vspace{0.1cm} 
    
    \tsnefigurecifar{resnet18_cifar10}{adaptive_patch}{Adaptive-Patch}\hfill
    \tsnefigurecifar{resnet18_cifar10}{adaptive_blend}{Adaptive-Blend}\hfill
    \tsnefigurecifar[0.02]{resnet18_cifar10}{bpp}{Bpp}
    
    \vspace{0.1cm} 
    
    \tsnefigurecifar{resnet18_cifar10}{dfst}{DFST}\hfill
    \tsnefigurecifar{resnet18_cifar10}{narcissus}{Narcissus}\hfill
    \tsnefigurecifar{resnet18_cifar10}{grond}{Grond}
    
    \vspace{0.1cm} 
    
    \begin{minipage}{0.32\textwidth}
        \centering \small PR=0.05 \hspace{0.8cm} PR=0.003
    \end{minipage}\hfill
    \begin{minipage}{0.32\textwidth}
        \centering \small PR=0.05 \hspace{0.8cm} PR=0.003
    \end{minipage}\hfill
    \begin{minipage}{0.32\textwidth}
        \centering \small PR=0.05 \hspace{0.8cm} PR=0.003$^*$
    \end{minipage}
    
    \vspace{-0.2cm} 
    \caption{t-SNE feature space embeddings of backdoored models on the CIFAR-10 training dataset, poisoned with the specified poisoning rate (PR). Benign samples are colored \textcolor{blue}{blue}, while poisoned samples are colored \textcolor{red}{red}. *WaNet and Bpp use PR=0.02.}
    \label{fig:cifar-tsne}
\end{figure*}
\section{Experimental Settings}
\label{sec:experimental_settings}

We evaluate footprints using the metrics in Table~\ref{tab:metrics} across the datasets, architectures, and attacks described below.

\paragraph{Datasets}
We use CIFAR-10/100~\cite{krizhevsky2009cifar} (50k/10k, RGB, $32{\times}32$), Imagenette~\cite{repoImagenette} (10 classes; we center-crop to square and downscale to $80{\times}80$), and Tiny-ImageNet~\cite{le2015tiny} (200 classes, native $64{\times}64$). Unless specified otherwise, we follow BackdoorBench data processing for each dataset. 

\paragraph{Architectures}
We use ResNet-18~\cite{he2015resnets}, VGG-16~\cite{simonyan2015vgg}, and ViT-Small~\cite{alexey2021vit} for conducting experiments. Implementations are aligned across attacks (e.g., adaptive average pooling for non-$32{\times}32$ inputs) to ensure comparable feature dimensions and training behavior.

\paragraph{Implementations}
We use official repositories for Adap-Patch/Blend, DFST, Grond, DFBA, and BackdoorBench~\cite{baoyuan2022bb} for BadNets, Blend, WaNet, BppAttack, and benign baselines; Narcissus is ported into BackdoorBench. Minor fixes standardize normalization, batching, and trigger I/O across attacks. The attack-specific settings we adjust or standardize, and the implementation of each footprint metric ($\ell_p$ norms, MSE, PSNR, SSIM~\cite{wang2004ssim}, LPIPS~\cite{zhang2018lpips}, IS~\cite{salimans2016is}, pHash, SAM~\cite{yuhas1992sam}, SS~\cite{rousseeuw1987silhouette}, DSWD~\cite{doan2021dswd}, and UCLC and TAC~\cite{zheng2022clp}), are detailed in the supplementary material.

\subsection{Training Settings}
\label{sec:exp_settings:train_settings}

\textbf{Optimizer and schedule.} All models train from scratch with SGD (momentum 0.9, weight decay $5{\times}10^{-4}$) and MultiStepLR (milestones at epochs 100 and 150 with $\gamma{=}0.1$; total 200 epochs). Batch size is 100 (CIFAR-10/100), 20 (Imagenette), and 64 (Tiny-ImageNet), chosen to (i) fit memory and (ii) preserve exact poisoned counts for $5\%$ PR in batch-wise poisoning (see below). These choices follow settings from related work.

\textbf{Targets and poisoning rates.} We use all-to-one attacks targeting class index 0 in each dataset. Poisoning rates (PR) are $5\%$ and $0.3\%$, with a few exceptions. WaNet and BppAttack on CIFAR-10 use $2\%$ instead of $0.3\%$, as their imperceptible triggers are difficult for backdoored models to learn at low poisoning rates. Clean-label attacks (Narcissus, Grond) use $0.7\%$ on CIFAR-100 and $0.4\%$ on Tiny ImageNet instead of $5\%$, since only target-class samples (which make up $1\%$ of CIFAR-100 and $0.5\%$ of Tiny ImageNet) can be poisoned. For batch-wise poisoning (WaNet, BppAttack, DFST), we adopt BackdoorBench’s rounding fix: if $\text{PR}\times\text{batch\_size} < 1$, a single poisoned example is inserted into a fraction of batches so that the dataset-level fraction matches PR in expectation.

\textbf{Normalization.} We normalize inputs during training and evaluation with the per-channel mean and standard deviation of each dataset, which are listed in the supplementary material; when a trigger generator operates on unnormalized pixels (e.g., UPGD in Grond), normalization is applied immediately before model forward passes to keep $l_\infty$ constraints on the unnormalized perturbation well-defined.

\subsection{Additional Cross-Architecture Analysis}
\label{app:architecture_ranking}

\begin{figure}
    \centering
    \includegraphics[width=0.85\linewidth]{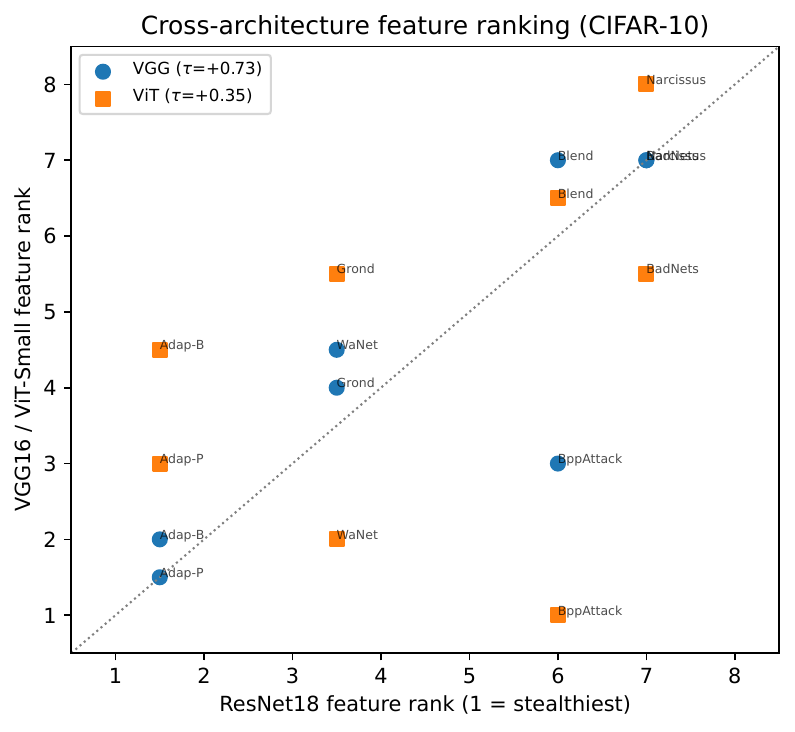}
    \caption{Combined feature-space footprint rank of each attack on CIFAR-10 for ResNet18 versus VGG16 and ViT-Small. Rank 1 denotes the smallest measured feature-space footprint. VGG16 broadly follows the
ResNet18 ordering, whereas ViT-Small diverges, reflecting differences in representation geometry and the low ASR of several attacks on
ViT-Small.}
\label{fig:architecture_feature_ranks}
\end{figure}

Figure~\ref{fig:architecture_feature_ranks} compares the combined
feature-space rankings across architectures.

\section{Threat Model and Access Assumptions}
\label{app:threat}
\label{app:access}

\begin{figure}[t]
\centering
\definecolor{spinput}{HTML}{2A78D6}
\definecolor{spfeature}{HTML}{EB6834}
\definecolor{spparam}{HTML}{1BAF7A}
\begin{tikzpicture}[x=1pt, y=1pt, font=\scriptsize, >=stealth,
    box/.style={draw=black!55, rounded corners=1.5pt, inner sep=2.5pt, align=center}]
  \node[box, fill=black!5] (att) at (122, 84) {\textbf{Attacker:} poisons data, controls training, or edits weights};
  \draw[->, black!70] (att.south) -- (122, 66);
  \draw[black!55, rounded corners=2pt] (64, 14) rectangle (196, 66);
  \node at (130, 59.5) {backdoored model};
  \foreach \xx in {80, 97, 114}
    \draw[fill=spparam!20, draw=spparam!75!black] (\xx-4, 26) rectangle (\xx+4, 51);
  \draw[->, black!70] (118, 38.5) -- (151, 38.5);
  \node[box, fill=spfeature!20, draw=spfeature!75!black] (h) at (165, 38.5) {$h$};
  \node at (97, 19.5) {weights $\theta$};
  \node at (165, 19.5) {features};
  \node[box, fill=spinput!15, draw=spinput!75!black] (x) at (26, 50) {clean $x$};
  \node[box, fill=spinput!15, draw=spinput!75!black] (tx) at (26, 27) {triggered $T(x)$};
  \draw[->, black!70] (x.east) -- (64, 50);
  \draw[->, black!70] (tx.east) -- (64, 27);
  \node[anchor=west] (y) at (204, 50) {true $y$};
  \node[anchor=west] (t) at (204, 27) {target $t$};
  \draw[->, black!70] (196, 50) -- (y.west);
  \draw[->, black!70] (196, 27) -- (t.west);
  \draw[spinput, line width=1.4pt] (1, 6) -- (51, 6);
  \draw[spparam, line width=1.4pt] (72, 6) -- (122, 6);
  \draw[spfeature, line width=1.4pt] (146, 6) -- (184, 6);
  \node at (26, -1) {\textbf{Input space}};
  \node at (97, -1) {\textbf{Parameter space}};
  \node at (165, -1) {\textbf{Feature space}};
  \node at (26, -9.5) {$\ell_p$, SSIM, LPIPS, \dots};
  \node at (97, -9.5) {UCLC, TAC, TUP};
  \node at (165, -9.5) {SS, DSWD, CDBI};
\end{tikzpicture}
\caption{Threat model. The backdoored model predicts the target class $t$ for triggered inputs. We measure the footprint of the backdoor in the input space (clean versus triggered input), the parameter space (weights), and the feature space (penultimate-layer features $h$).}
\label{fig:threat_model}
\end{figure}
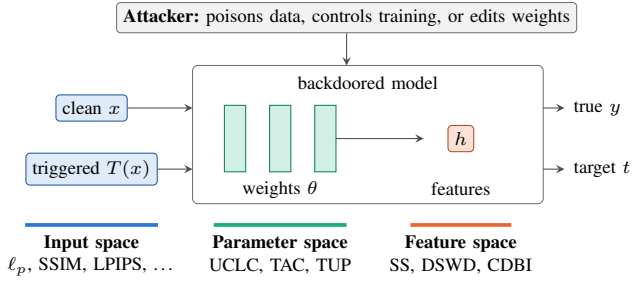

\paragraph{Threat model}
In machine learning security, a backdoor is a hidden rule that an attacker places in a model: the backdoored model classifies clean inputs correctly, but it predicts an attacker-chosen target class whenever the input contains the trigger. The attacker wants a high ASR, a benign accuracy close to that of a clean model, and a backdoor that is hard to notice. The evaluated attacks differ in what the attacker controls (Table~\ref{tab:taxonomy}): five of them only poison part of the training data, four additionally control the training process, and DFBA modifies a trained model without any training data. Figure~\ref{fig:threat_model} shows where a backdoor leaves a footprint. In the input space, the trigger changes the image. In the feature space, the model can represent triggered inputs differently from clean inputs. In the parameter space, the backdoor is stored in the weights and can make some channels unusually sensitive to the trigger. These are three views of the same backdoor, not three different attacks.

\paragraph{Attack taxonomy}
Table~\ref{tab:taxonomy} lists the ten attacks by family, as in Section~\ref{sec:benchmark_results}, together with the mean footprint of each attack in each space. Attacks with a small footprint in one space usually have a larger footprint in another.

\definecolor{fpcell}{HTML}{2A78D6}
\begin{table}[t]
\centering
\scriptsize
\caption{The evaluated attacks. Access: the attacker poisons part of the training data (data), controls the training process (training), or modifies a trained model (model); Narcissus and Grond keep the correct labels of the poisoned samples. The last three columns give the mean footprint of each attack over all its configurations, on the scale of Table~\ref{tab:asr_sensitivity} (within-dataset percentile, 0 = smallest); darker cells mark larger footprints. DFBA's feature-space value uses DSWD only.}
\label{tab:taxonomy}
\setlength{\tabcolsep}{3pt}
\begin{tabular}{@{}llllccc}
\toprule
Family & Attack & Access & Trigger & Input & Feature & Param. \\
\midrule
\multirow{2}{*}{Patch/blend} & BadNets & data & patch & \cellcolor{fpcell!27}0.37 & \cellcolor{fpcell!44}0.70 & \cellcolor{fpcell!44}0.69 \\
 & Blend & data & blended image & \cellcolor{fpcell!40}0.62 & \cellcolor{fpcell!39}0.59 & \cellcolor{fpcell!31}0.45 \\
\midrule
\multirow{2}{*}{Imperceptible} & WaNet & training & warping & \cellcolor{fpcell!24}0.31 & \cellcolor{fpcell!37}0.56 & \cellcolor{fpcell!44}0.69 \\
 & BppAttack & training & quantization & \cellcolor{fpcell!16}0.15 & \cellcolor{fpcell!43}0.67 & \cellcolor{fpcell!44}0.70 \\
\midrule
\multirow{2}{*}{Adaptive} & Adap-Patch & data & patches & \cellcolor{fpcell!38}0.57 & \cellcolor{fpcell!20}0.23 & \cellcolor{fpcell!35}0.52 \\
 & Adap-Blend & data & blended image & \cellcolor{fpcell!44}0.69 & \cellcolor{fpcell!26}0.34 & \cellcolor{fpcell!29}0.40 \\
\midrule
\multirow{2}{*}{Clean-label} & Narcissus & data & optimized & \cellcolor{fpcell!54}0.89 & \cellcolor{fpcell!27}0.37 & \cellcolor{fpcell!28}0.38 \\
 & Grond & training & optimized & \cellcolor{fpcell!30}0.42 & \cellcolor{fpcell!25}0.33 & \cellcolor{fpcell!18}0.19 \\
\midrule
\multirow{2}{*}{Supply-chain} & DFST & training & style filter & \cellcolor{fpcell!52}0.86 & \cellcolor{fpcell!53}0.87 & \cellcolor{fpcell!35}0.52 \\
 & DFBA & model & patch & \cellcolor{fpcell!22}0.27 & \cellcolor{fpcell!60}1.00 & \cellcolor{fpcell!55}0.91 \\
\bottomrule
\end{tabular}
\end{table}

\paragraph{Access requirements of the metrics}
Our footprint metrics are intended for offline evaluation, such as benchmarking attacks or auditing a model with full access. They are not detection methods for a defender who knows neither the trigger nor the poisoned data (see also the limitations in Section~\ref{sec:guidance_limitations}). Table~\ref{tab:access} summarizes what each metric requires. Only UCLC can be computed from the model weights alone. All other metrics need clean data and either the trigger or the poisoned training samples, and the feature-space metrics, TAC, and TUP also need the activations of the backdoored model.

\begin{table}[t]
\centering
\footnotesize
\caption{Information required by each metric. Poisoned samples: the poisoned training set and which of its samples are poisoned. Model: A~=~activations and W~=~weights of the backdoored model; R~=~a fixed pretrained network (AlexNet for LPIPS, Inception~v3 for IS). Benign model: a clean model trained on the same data.}
\label{tab:access}
\setlength{\tabcolsep}{3pt}
\begin{tabular}{@{}L{0.30\linewidth}ccccc@{}}
\toprule
Metrics & \begin{tabular}{@{}c@{}}Clean\\data\end{tabular} & Trigger & \begin{tabular}{@{}c@{}}Poisoned\\samples\end{tabular} & Model & \begin{tabular}{@{}c@{}}Benign\\model\end{tabular} \\
\midrule
$\ell_1$, $\ell_2$, $\ell_\infty$, MSE, PSNR, SSIM, pHash, SAM & \checkmark & \checkmark & -- & -- & -- \\
LPIPS, IS  & \checkmark & \checkmark & -- & R & -- \\
\midrule
SS, CDBI   & \checkmark & -- & \checkmark & A & -- \\
DSWD       & \checkmark & \checkmark & -- & A, W & -- \\
\midrule
UCLC       & -- & -- & -- & W & -- \\
TAC        & \checkmark & \checkmark & -- & A & -- \\
TUP        & \checkmark & \checkmark & -- & A, W & \checkmark \\
\bottomrule
\end{tabular}
\end{table}

\section{Robustness Analysis}
\label{app:robust}

This appendix reports additional experiments that test the robustness of our main results. The complete results of these experiments are provided in the supplementary material.

\subsection{Training Seeds and Target Classes}
\label{app:robust:seeds}

In our main experiments, each configuration is trained once, with target class~0. To test whether the measurements depend on these choices, we retrain BadNets, Blend, WaNet, and BppAttack on CIFAR-10 (ResNet18, 5\% poisoning rate) with four additional random seeds and with two other target classes (1 and 2). The benign reference models used by TUP are retrained with the same seeds. For each model, we recompute the six metrics that depend on the trained model: SS, CDBI, DSWD, UCLC, TAC, and TUP.

For BadNets, Blend, and BppAttack, benign accuracy and ASR vary by less than 0.5 percentage points across seeds. WaNet is the exception: its ASR ranges from 63.6\% to 99.8\%, and its footprints are smaller in the two runs with a weaker backdoor (for example, a TUP of 5.5 and 6.6 instead of 8.9--9.6). This agrees with the effect of attack success on footprint size discussed in Section~\ref{sec:asr_effectiveness}. Changing the target class changes benign accuracy and ASR by less than one percentage point for all four attacks.

Table~\ref{tab:robust:seeds} summarizes the stability of the metrics. Across seeds, the feature-space metrics and TUP vary only slightly (median coefficient of variation of 1.4--6.9\%), while UCLC and TAC vary by about 17\%. TUP ranks the four attacks in the same order for every seed and every target class, and SS and DSWD also keep the same order across target classes. CDBI and TAC swap one pair of neighboring attacks, whereas the UCLC ranking changes considerably at target class~2. The lower rank correlations of the feature-space metrics across seeds are mainly due to close values: for instance, the DSWD values of BadNets, Blend, and BppAttack all lie between 1.90 and 2.11 across the five seeds, so small changes can swap their order. In summary, retraining changes the feature-space metrics and TUP only slightly as long as the backdoor is successfully implanted, while single UCLC and TAC values should be interpreted with care.

\begin{table}[t]
\centering
\footnotesize
\caption{Stability of the model-dependent metrics over five training seeds and three target classes (BadNets, Blend, WaNet, and BppAttack on CIFAR-10, ResNet18). CV is the median coefficient of variation over seeds; $\tau$ is Kendall's rank correlation between the attack orderings, averaged over all pairs of seeds or target classes.}
\label{tab:robust:seeds}
\begin{tabular}{lccc}
\toprule
Metric & CV (seeds) & $\tau$ (seeds) & $\tau$ (targets) \\
\midrule
SS   & 2.9\%  & 0.60    & 1.00 \\
CDBI & 6.0\%  & 0.27    & 0.78 \\
DSWD & 1.4\%  & 0.33    & 1.00 \\
\midrule
UCLC & 16.6\% & $-0.07$ & 0.33 \\
TAC  & 16.5\% & 0.67    & 0.78 \\
TUP  & 6.9\%  & 1.00    & 1.00 \\
\bottomrule
\end{tabular}
\end{table}

\subsection{Uncertainty of the Metric Agreement}
\label{app:robust:bootstrap}

The agreement analysis in Section~\ref{sec:metric_disagreement} treats the 76 configurations as independent observations, and so do the significance marks in Figure~\ref{fig:metric_correlation}. However, configurations of the same attack share the trigger and the implementation, so they are not independent. We therefore compute 95\% confidence intervals with a bootstrap that resamples whole attacks: each of the 2,000 resamples draws ten attacks with replacement and keeps all configurations of every drawn attack. We summarize the main result of Section~\ref{sec:metric_disagreement} by the gap between the average of the three within-space blocks and the average of the three across-space blocks of Table~\ref{tab:metric_correlation_blocks}, and we repeat the analysis on the 61 configurations with an ASR of at least 60\%.

Table~\ref{tab:robust:bootstrap} shows the results. The gap is $+0.50$ with an interval of $[0.28, 0.62]$, and it stays almost the same when only successful backdoors are used ($+0.50$, $[0.27, 0.64]$). The strongest cross-space relation in Figure~\ref{fig:metric_correlation}, between $\ell_1$ and TUP, also remains clearly negative, with an interval of $[-0.80, -0.47]$. Two averages are less certain: the intervals of the parameter--parameter and input--parameter blocks include zero, although they would exclude zero if configurations were resampled independently. Finally, removing the low-ASR configurations makes the input--parameter disagreement stronger ($-0.24$ to $-0.32$), so this disagreement is not caused by failed backdoors.

Some metrics in our suite are derived from others: PSNR is a monotone function of MSE, CDBI uses the same embedding as SS, and TUP combines TAC and UCLC. Such pairs can inflate the agreement within a space. Without PSNR, CDBI, and TUP, the gap decreases from $+0.50$ to $+0.35$, but its interval of $[0.09, 0.52]$ still excludes zero, also when only successful backdoors are used. The largest change is in the parameter space, where UCLC and TAC alone agree only weakly ($+0.05$, pooled over datasets).

\begin{table}[t]
\centering
\footnotesize
\caption{Pooled block averages of Table~\ref{tab:metric_correlation_blocks} with 95\% confidence intervals from a bootstrap over attacks (2,000 resamples), for all configurations and for those with ASR $\geq 60\%$.}
\label{tab:robust:bootstrap}
\setlength{\tabcolsep}{4pt}
\begin{tabular}{lcc}
\toprule
Metric group & All ($N{=}76$) & ASR $\geq 60\%$ ($N{=}61$) \\
\midrule
Input--Input         & $+0.66$ [$0.43, 0.80$]  & $+0.69$ [$0.43, 0.82$] \\
Feature--Feature     & $+0.57$ [$0.42, 0.65$]  & $+0.44$ [$0.29, 0.54$] \\
Parameter--Parameter & $+0.27$ [$-0.04, 0.52$] & $+0.24$ [$-0.06, 0.50$] \\
Input--Feature       & $-0.11$ [$-0.41, 0.24$] & $-0.11$ [$-0.43, 0.24$] \\
Input--Parameter     & $-0.24$ [$-0.41, 0.07$] & $-0.32$ [$-0.51, 0.05$] \\
Feature--Parameter   & $+0.35$ [$0.20, 0.49$]  & $+0.29$ [$0.13, 0.43$] \\
\midrule
Within $-$ across    & $+0.50$ [$0.28, 0.62$]  & $+0.50$ [$0.27, 0.64$] \\
\bottomrule
\end{tabular}
\end{table}

\subsection{Remaining Configurations after ASR Filtering}
\label{app:robust:asr}

Table~\ref{tab:robust:asr} lists how many configurations of each attack remain after the ASR filters of Section~\ref{sec:asr_effectiveness}, which keep 61 of the 76 configurations at 60\% and 51 at 80\%. Most removed configurations belong to Adap-Patch, Grond, and BppAttack. Adap-Patch and Grond, which have the smallest feature-space and parameter-space footprints, respectively, keep only one configuration each at 80\%.

\begin{table}[t]
\centering
\footnotesize
\caption{Configurations per attack that remain after removing those with ASR below 60\% or 80\%. Each attack has eight configurations, except DFBA with four.}
\label{tab:robust:asr}
\setlength{\tabcolsep}{4pt}
\begin{tabular}{lcc@{\hspace{12pt}}lcc}
\toprule
Attack & $\geq$60\% & $\geq$80\% & Attack & $\geq$60\% & $\geq$80\% \\
\midrule
BadNets    & 8 & 8 & Adap-Blend & 6 & 6 \\
Blend      & 8 & 7 & DFST       & 8 & 8 \\
WaNet      & 8 & 7 & Narcissus  & 6 & 5 \\
BppAttack  & 6 & 4 & Grond      & 5 & 1 \\
Adap-Patch & 2 & 1 & DFBA       & 4 & 4 \\
\bottomrule
\end{tabular}
\end{table}

\subsection{Re-runs of the Per-Space Winners}
\label{app:robust:winners}

Adap-Patch and Grond have the smallest feature-space and parameter-space footprints, respectively, but each keeps only one configuration under the 80\% ASR filter (Appendix~\ref{app:robust:asr}). To test whether these results depend on one particular training run, we retrain both attacks in this configuration (CIFAR-10, ResNet18, with poisoning rates of 0.3\% for Adap-Patch and 5\% for Grond) 14 more times: four times with target class 0 and five times each with target classes 1 and 2. Together with the original model, this gives 15 runs per attack, which we compare with the 28 models of Appendix~\ref{app:robust:seeds}.

All 15 Grond runs and 13 of the 15 Adap-Patch runs reach an ASR of at least 80\%, and the weakest Adap-Patch run still reaches 71.5\%. As Table~\ref{tab:robust:winners} shows, Adap-Patch has a smaller feature-space footprint than all 28 comparison models under SS and CDBI in every run, and Grond has a smaller parameter-space footprint under UCLC and TUP in every run. For DSWD and TAC, the values of the two attacks overlap with those of the comparison models. These small footprints are therefore not the result of a single training run or of the choice of target class.

\begin{table}[t]
\centering
\footnotesize
\caption{Range of each metric over the 15 runs of Adap-Patch and Grond (three target classes), and the best value among the 28 comparison models of Appendix~\ref{app:robust:seeds}. Arrows show the direction of a smaller footprint.}
\label{tab:robust:winners}
\begin{tabular}{llcc}
\toprule
Metric & Attack & Runs (range) & Others (best) \\
\midrule
SS $\downarrow$   & Adap-Patch & 0.02--0.20  & 0.37 \\
CDBI $\uparrow$   & Adap-Patch & 1.42--10.76 & 0.73 \\
UCLC $\downarrow$ & Grond      & 3.44--4.60  & 6.04 \\
TUP $\downarrow$  & Grond      & 1.68--1.87  & 3.61 \\
\bottomrule
\end{tabular}
\end{table}

\subsection{Sensitivity of CDBI to the Embedding}
\label{app:robust:cdbi}

We compute SS and CDBI on a two-dimensional t-SNE embedding of the penultimate-layer features with perplexity 30. The values in Tables~\ref{tab:results_feature_resnet18} and~\ref{tab:results_newmetric_resnet18} come from a single embedding computed before we fixed the random seed of t-SNE. Our released code and the experiments in Appendices~\ref{app:robust:seeds} and~\ref{app:robust:winners} use a fixed seed (random state 0), so their SS and CDBI values are not directly comparable with these two tables.

Because t-SNE is random and does not preserve distances, the CDBI ranking may depend on the embedding. To test this, we recompute CDBI for 54 configurations (18 each on CIFAR-10, CIFAR-100, and Imagenette) in 18 representations of the same features: the raw 512-dimensional features, PCA with 50 and 2 dimensions, and t-SNE with perplexities 15, 30, and 50, each with five random seeds. We compare the resulting attack rankings with Spearman correlation.

Table~\ref{tab:robust:cdbi} shows that the CDBI ranking hardly depends on the t-SNE seed (mean correlation between seeds of at least 0.93 at each perplexity) and agrees well with the ranking in Table~\ref{tab:results_newmetric_resnet18}. However, the CDBI values change by 12--14\% across t-SNE runs, and the ranking on the t-SNE embedding agrees only moderately with the ranking on the raw features. CDBI should therefore be used to compare attacks within one stated embedding, not as an absolute value.

\begin{table}[t]
\centering
\footnotesize
\caption{Agreement between CDBI rankings in different representations (Spearman correlation) and variation of the CDBI values across the 15 t-SNE runs (median coefficient of variation). The comparison with Table~\ref{tab:results_newmetric_resnet18} uses the configurations that appear in that table (16 on CIFAR-10).}
\label{tab:robust:cdbi}
\setlength{\tabcolsep}{4pt}
\begin{tabular}{lccc}
\toprule
 & C10 & C100 & Imagenette \\
\midrule
t-SNE seeds (lowest mean)            & 0.98  & 0.93  & 0.97 \\
All representations vs.\ Table~\ref{tab:results_newmetric_resnet18} & 0.92 & 0.89 & 0.93 \\
t-SNE vs.\ raw features              & 0.52  & 0.53  & 0.84 \\
Variation of values                  & 12\%  & 14\%  & 12\% \\
\bottomrule
\end{tabular}
\end{table}

\subsection{Attack Success versus Footprint}
\label{app:robust:pareto}

\begin{figure}[t]
\centering
\includegraphics[width=\linewidth]{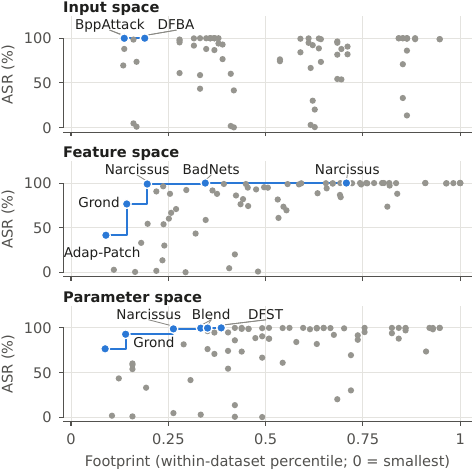}
\caption{ASR versus footprint of all 76 configurations in each space (footprint: mean within-dataset percentile of the metrics of the space, 0 = smallest). Blue points are configurations that no other configuration dominates (at least the same ASR and at most the same footprint, and better in one of the two).}
\label{fig:robust:pareto}
\end{figure}

Figure~\ref{fig:robust:pareto} plots the ASR of each configuration against its footprint in each space. As in Section~\ref{sec:asr_effectiveness}, the footprint in a space is the mean within-dataset percentile of the metrics of that space. The blue points mark the configurations that no other configuration dominates, i.e., no other configuration has at least the same ASR and at most the same footprint while being better in one of the two. These are the best trade-offs between attack success and footprint among the evaluated configurations.

In the input space, there is almost no trade-off: BppAttack and DFBA have the smallest footprints and an ASR close to 100\%. In the feature and parameter spaces, the smallest footprints come with a lower ASR. The configuration with the smallest feature-space footprint, Adap-Patch on Tiny-ImageNet, has an ASR of only 41.6\%, and the two blue points with the smallest parameter-space footprints, both Grond on CIFAR-10, reach 76.6\% and 92.9\%. Configurations with an ASR close to 100\% leave larger footprints in these two spaces, and we observe the same pattern when we compute the front for each dataset separately. Small feature-space and parameter-space footprints should therefore always be reported together with the ASR.

\FloatBarrier
\newpage

\section{Ethics Considerations}
\label{app:ethics}

This work studies how existing backdoor attacks leave measurable footprints across input, feature, and parameter space. Backdoor research is inherently dual-use, since improved evaluation may also reveal limitations of current defenses. However, we do not introduce a new attack, increase attack effectiveness, or propose a new method for evading detection. Our goal is to improve defensive evaluation by showing where commonly used stealthiness metrics disagree and why claims based on a single metric, visualization, or defense can be misleading.

All experiments are conducted offline on publicly available image-classification datasets using models trained or modified by us for research purposes. The work involves no human subjects, personally identifiable information, user data, or interaction with deployed production systems. We release code, configurations, and reproducibility artifacts to support independent verification and defensive research. The artifact is intended for use only on systems that researchers own or are authorized to evaluate.

\newpage 

\section{Meta-Review}

The following meta-review was prepared by the program committee for the 2027
IEEE Symposium on Security and Privacy (S\&P) as part of the review process as
detailed in the call for papers.

\subsection{Summary}
This paper presents a measurement study of backdoor stealthiness in image classification. The paper organizes existing measurements into three observation spaces---input, feature, and parameter---and systematically measures the footprints of ten backdoor attacks across these spaces. The evaluation shows that measurements can disagree both within and across observation spaces, suggesting that evidence of stealthiness in one space does not necessarily imply stealthiness in another.

\subsection{Scientific Contributions}
\begin{itemize}
\item Independent Confirmation of Important Results with Limited Prior Research
\item Provides a Valuable Step Forward in an Established Field
\item Provides New Experimental Insight into an Important Security Property
\end{itemize}

\subsection{Reasons for Acceptance}
\begin{enumerate}
\item The paper provides a systematic measurement study of backdoor stealthiness, a widely used but inconsistently quantified property in the backdoor literature. Its three-space framework provides a useful structure for comparing existing notions and metrics of stealthiness.
\item The paper demonstrates that stealthiness measurements can disagree substantially across observation spaces, and that no evaluated attack is uniformly stealthy across all three spaces. These findings provide useful guidance for evaluating future backdoor attacks and stealthiness claims.
\item The evaluation is extensive, covering ten attacks, multiple datasets and architectures, multiple stealthiness metrics, and analyses across seeds, target classes, and attack-success thresholds. The resulting methodology and empirical findings provide a useful foundation for more systematic evaluation of backdoor stealthiness.
\end{enumerate}



\end{document}